# On a multivariate extension for Copula-based Conditional Value at Risk

Andres Mauricio Molina Barreto [1*]

1* Faculty of Economics, Keio University, 2-15-45 Mita, Minato City, 108-8345, Tokyo, Japan.

Corresponding author(s). E-mail(s): molina.mauricio@keio.jp;

## Abstract

Copula-based Conditional Value at Risk (CCVaR) is defined as an alternative version of the classical Conditional Value at Risk (CVaR) for multivariate random vectors intended to be real-valued. We aim to generalize **CCVaR** to several dimensions ( $d \geq 2$ ) when the dependence structure is given by an Archimedean copula. While previous research focused on the bivariate case, leaving the multivariate version unexplored, an almost closed-form expression for CCVaR under an Archimedean copula is derived. The conditions under which this risk measure satisfies coherence are then examined. Finally, numerical experiments based on real data are conducted to estimate CCVaR, and the results are compared with classical measures of Value at Risk (VaR) and Conditional Value at Risk (CVaR).

# 1 Introduction

Because of its simplicity, Value at Risk has long been the benchmark for assessing the risk of financial instruments. Value at Risk (VaR) is defined as the worst loss given a confidence level $\beta (0 \leq \beta < 1)$. However, VaR has been widely criticized for its lack of sub-additivity. To overcome this, Conditional Value at Risk (CVaR) was introduced; it is defined as the expected loss given that the loss exceeds the VaR at level $\beta$. There is an extensive literature on VaR and CVaR estimation; notable contributions include Duffie and Pan (1997), Artzner, Delbaen, Eber, and Heath (1999), and Rockafellar and Uryasev (2000), as well as more recent developments such as those in McNeil, Frey, and Embrechts (2015). Some regulators use CVaR as an internal model, as recommended by Basel III (Bank for International Settlements, 2012).

Various terms are used in the literature to describe similar measures to CVaR, such as Tail Value at Risk (TVaR), Expected Shortfall (ES), and Average Value at Risk (AVaR). In this paper, we will use the term Conditional Value at Risk (CVaR), assuming that the variable $X$ is continuous. As noted, CVaR is a coherent risk measure, while VaR generally is not particularly in the presence of heavy-tailed distributions.

One classical approach to CVaR estimation is to consider the total weighted loss of the portfolio and use methods such as historical quantile estimation or fitting a parametric loss distribution. Additional techniques include nonparametric methods that avoid distributional assumptions (Embrechts, Klüppelberg, & Mikosch, 2008; Scaillet, 2023; Yamai & Yoshiba, 2005), as well as estimation via GARCH models to account for volatility clustering (Christoffersen, 1998). Engle and Manganelli (2004) proposed a conditional autoregressive Value at Risk (CAViaR) measure that specifies the evolution of the quantile over time using an autoregressive process.

In multivariate settings, several extensions of VaR and CVaR have been proposed to account for the joint behavior of multiple risk factors. For example, Prékopa (2012) introduced a multivariate VaR (MVaR) defined as the quantile set of a multivariate distribution, while multivariate CVaR (MCVaR) has been developed alongside portfolio optimization models (see, e.g., Lee and Prékopa, 2013).

An alternative approach to risk estimation involves the use of copulas. The copula-based CVaR (denoted here as CCVaR), introduced by Krzemienowski and Szymczyk (2016), offers a flexible method for modeling dependencies, though it can be computationally demanding even in the independence copula case. Other multivariate generalizations of VaR and CVaR, such as the Multivariate Lower-Orthant and Upper-Orthant VaR introduced by Cousin and Di Bernardino (2013), define risk over specific level sets where the risk contributions of different factors are heterogeneous and cannot be simply aggregated (Cousin & Di Bernardino, 2014). Later, Hürlimann (2014) further investigated properties of these vector-valued risk measures.

Copulas have also been used to model the dependence structure among several random variables. A copula is a function whose margins are uniformly distributed on [0,1], and by Sklar's theorem, the joint distribution function can be expressed in terms of its marginal distributions and a copula (Nelsen, 2006; Sklar, 1959). In finance, copulas are widely applied in portfolio management, as they can flexibly capture the dependence structure among assets (Cherubini, 2004). Among copulas, the Archimedean family has several desirable properties, including ease of estimation and simulation, because these copulas are generated by a single generator function that is (up to a constant) the Laplace-Stieltjes transform of a probability distribution.

A study on the implementation of a more straightforward alternative formula for CCVaR was developed in Molina Barreto and Ishimura (2023), which is a measure with real values including the effects of Archimedean copulas and analyzing its relationship with MCVaR. However, their study focuses on examples with bivariate copulas, leaving the general case of higher dimensions unexplored. Also, the number of copulas included in their study seems to be small, leaving more room to study the Archimedean copulas most commonly used in this field.

Although Archimedean copulas are exchangeable (i.e., symmetric) and thus offer a limited dependence structure, this assumption may be acceptable for portfolios comprising risk factors from similar industries. Moreover, this exchangeability permits to derive an almost closed-form expression for the copula-based Conditional Value at Risk, which greatly facilitates theoretical analysis, numerical implementation, and calibration in higher dimensions. While one might alternatively consider ordering the risk factors in decreasing order of dependence before computing the copula function, such an approach would introduce additional complexity into the risk measure. By instead relying on the multivariate extension of the Kendall function, the presented method captures the essential tail dependence structure in a straightforward manner.

In Section 2, the preliminaries of VaR, CVaR, and Archimedean copulas are reviewed. In Section 3, the main results and properties of the proposed CCVaR are discussed. In subsection 3.2, empirical applications to real data from machinery stocks listed on the Nikkei 225 index are presented. Finally, in Section 4, findings and outline potential avenues for future research are discussed.

# 2 Theoretical Basis

## 2.1 Value at Risk and Conditional Value at Risk

Value at Risk is typically defined for a single random variable and represents the worst possible losses within a given pre-specified confidence level. Consider a probability space ( $\Omega, \mathcal{F}, P$ ). Let $X$ be a real-valued random variable, and for simplicity, let us suppose it is continuous. If $F_X(x) = \mathbb{P}(X \leq x)$ is the cumulative distribution function (cdf) of the variable $X$, the VaR at the confidence level $\beta (0 \leq \beta < 1)$ can be defined as

$$\text{VaR}_\beta(X) := F_X^{(-1)}(\beta) = \inf\{u \mid F_X(u) \geq \beta\}.$$

where $F_X^{(-1)}(\cdot)$ is also known as its quantile function. However, VaR is not a coherent risk measure in the sense of Artzner et al. (1999). To overcome this, the Conditional Value at Risk is used (CVaR) was introduced. This is formulated as the expected value of the losses given that the actual loss exceeds the VaR at level $\beta$ (Pflug, 2000). In the general setting, if $\text{VaR}_\beta(X) = F_X^{-1}(\beta)$ is the $\beta$-quantile, the Conditional Value at Risk $\text{CCVaR}_\beta(X)$ is defined as the solution of the following optimization problem:

$$\text{CVaR}_\beta(X) := \inf\left\{a + \frac{1}{1-\beta} E[X-a]^+ : a \in \mathbb{R}\right\}$$

where $[x]^+ = \max(x, 0)$. In particular, if $F_X$ is smooth, CVaR results in the conditional expectation of $X$, given that $X > \text{VaR}_\beta(X)$.

$$\text{CVaR}_\beta(X) = E[X \mid X > \text{VaR}_\beta(X)]$$

Alternatively, if $\beta$ is in tha range of $F_X$, the $(\text{CVaR}_\beta)$ can be formulated as:

$$\text{CVaR}_\beta(X) := \frac{1}{1-\beta} \int_\beta^1 \text{VaR}_t(X) dt = \frac{1}{1-\beta} \int_\beta^1 F_X^{-1}(t) dt$$
$$= \frac{1}{1-\beta} \int_{F_X^{(-1)}(\beta)}^\infty u \, dF_X(u)$$

A well-recognized fact about CVaR is that is a coherent risk measure in the sense of Artzner et al. (1999), while VaR fails to satisfy the sub-additivity property for general loss distributions. Also, CVaR gives more importance to losses that occur during extreme events than the usual notion of VaR.

Some contributions for extending the concept of VaR and CVaR for multivariate risk factors have been developed in recent years. For example, Prékopa (2012) defined a Multivariate Value at Risk (MVaR) as the quantile set of multivariate probability distribution, also known as $\beta$-efficient points. If **X** is a continuous random vector of dimension $d$ with strictly increasing joint distribution $H(\mathbf{X})$, then

$$\text{MVaR}_\beta(\mathbf{X}) = \{(x_1, \ldots, x_d) \mid H(x_1, \ldots, x_d) = \beta\}$$

The notation $D_\beta$ and $D_\beta^c$ represent the favorable and unfavorable sets, respectively, where

$$D_\beta = \bigcup_{x \in \text{MVaR}_\beta(\mathbf{X})} (x + \mathbb{R}_+^d)$$

Definition 1. The Multivariate Conditional Value at Risk (MCVaR) of the random vector **X** for a risk level $\beta$ is defined as:

$$\text{MCVaR}_\beta(\mathbf{X}) = E\left[\Psi(\mathbf{X}) \mid \mathbf{X} \in \overline{D_\beta^c}\right] \qquad (1)$$

where $\Psi$ is some $d$-variate function such that $E[\Psi(X)]$ exists. The set $\overline{D_\beta^C}$ designates the closure of $\overline{D_\beta^C}$.

If the function $\Psi(\cdot)$ is the convex sum of the components of $\mathbf{X}$ with weights given by $\lambda = (\lambda_1, \ldots, \lambda_d)^t$ and $\sum_{i=1}^d \lambda_i = 1$, then the MCVaR$_\beta$ of the corresponding portfolio is given by:

$$\text{MCVaR}_\beta(\mathbf{X}) = E[\lambda^t X \mid H(X) \geq \beta], \text{ where } 0 \leq \beta < 1 \tag{2}$$

Properties for MVaR and MCVaR were also studied (Lee & Prékopa, 2013; Prékopa, 2012), in which, the latest one satisfies the properties for a coherent risk measure under certain hypothesis.

## 2.2 Copula

The formal definition of a copula is stated as follows.

Definition 2. A copula $C^{(d)}$ of dimension $d (d \geq 2)$, is a multivariate function defined as $C: [0,1]^d \to [0,1]$ that follows the following properties (McNeil et al., 2015; Nelsen, 2006):

1. $C(u_1, \ldots, u_d)$ is increasing in each component $u_i$.
2. $C(1, \ldots, 1, u_i, 1, \ldots, 1) = u_i$ for all $i \in \{1, \ldots, d\}, u_i \in [0,1]$.
3. For all $(a_1, \ldots, a_d), (b_1, \ldots, b_d) \in [0,1]^d$ with $a_i \leq b_i$ we have

$$\sum_{i_1=1}^{2} \cdots \sum_{i_d=1}^{2} (-1)^{i_1 + \cdots + i_d} C(u_{1i_1}, \ldots, u_{di_d}) \geq 0$$

where $u_{j1} = a_j$ and $u_{j2} = b_j$ for all $j \in \{1, \ldots, d\}$.

The last property is known as the non-negative volume. It is easy to see that a copula is a continuous function by definition. For ease of notation in the formulas, the symbol $C$ is used to indicate the copula $C^{(d)}$ of dimension $d$, and the latter notation is used to highlight its dimension. These types of functions allow modeling the joint distribution function through their uniform margins. Let $\mathbf{X} = (X_1, \ldots, X_d)$ be a random vector with marginal distribution functions $F_i(x_i) = P(X_i \leq x_i)$ for $i = 1, \ldots, d$ and its joint probability function is given by:

$$H(x_1, \ldots, x_d) = P(X_1 \leq x_1, \ldots, X_d \leq x_d).$$

According to Sklar's theorem, the joint distribution function $H$ can be expressed with a copula $C^{(d)}$ as:

$$H(x_1, \ldots, x_d) = C(F_1(x_1), \ldots, F_d(x_d)).$$

This result is widely cited in the literature regarding copulas and can be considered central to this theory. It states that a copula function applied to transformed margins to uniform variables, results in the joint distribution with the dependence structure entirely given by the copula $C^{(d)}$ (Nelsen, 2006; Sklar, 1959).

One of the most famous copulas functions is from the family known as Archimedean Copulas, whose have several practical properties for modeling multivariate distributions. The ease of applying this class of copulas is given by the fact that they are completely determined by a generator function, that are inverses of Laplace-transforms of known probability distribution functions. These copulas offer flexibility in modeling dependence structures, and estimating their parameters is relatively simple compared to other families. For further details, we refer to McNeil and Nešlehová (2009). Let the generator $\varphi$ of a $d$-dimension copula $C^{(d)}$, a function that satisfies that $\varphi: [0,1] \to [0, \infty]$ is a continuous, strictly decreasing and convex function such that $\varphi(1) = 0$ and $\varphi(0) = \infty$. Its inverse function $\varphi^{-1}$ satisfies the same properties as monotonicity, but $\varphi^{-1}(0) = 1$ and $\varphi^{-1}(\infty) = 0$.

| Family | $\theta$ | $\varphi(t)$ | $C(u_1, \ldots, u_d)$ |
|---|---|---|---|
| Independence |  | $-\log t$ | $\prod_{i=1}^{d} u_i$ |
| Comonotonic |  |  | $\min\{u_1, \ldots, u_d\}$ |
| Clayton | $(0, \infty)$ | $t^{-\theta} - 1$ | $\left(1 - d + \sum_{i=1}^{d} u_i^{-\theta}\right)^{-1/\theta}$ |

| Frank | $(0, \infty)$ | $-\log\left(\frac{e^{-\theta t} - 1}{e^{-\theta} - 1}\right)$ | $-\frac{1}{\theta}\ln\left(1 + \frac{\prod_{i=1}^{d}(e^{-\theta u_i} - 1)}{(e^{-\theta} - 1)^{d-1}}\right)$ |
|---|---|---|---|
| Gumbel | $[1, \infty)$ | $(-\log t)^{\theta}$ | $\exp\left(-\left[\sum_{i=1}^{d}(-\ln u_i)^{\theta}\right]^{1/\theta}\right)$ |
| Joe | $[1, \infty)$ | $-\log\left(1 - (1-t)^{\theta}\right)$ | $1 - \left(1 - \prod_{i=1}^{d}\left[1 - (1 - u_i)^{\theta}\right]\right)^{1/\theta}$ |
| Ali-Mikhail-Haq | $[0,1)$ | $\log\left(\frac{1 - \theta(1-t)}{t}\right)$ | $(1 - \theta)\prod_{i=1}^{d} u_i / \left(\prod_{i=1}^{d}(1 - \theta(1 - u_i)) - \theta\prod_{i=1}^{d} u_i\right)$ |

Table 1 Some families of Archimedean copulas and generators

Definition 3. The function $C^{(d)}: [0,1]^d \to [0,1]$ is an Archimedean copula if can be expressed as:

$$C^{(d)}(u_1, \ldots, u_d) = \varphi^{-1}(\varphi(u_1) + \cdots + \varphi(u_d)) \quad (3)$$

and $\varphi^{-1}$ is $d$-monotone on $[0, \infty)$.

The inverse of the generator is said to be $d$-monotone, if it is differentiable up to order $d - 2$ and the derivatives satisfy:

$$(-1)^i \frac{d^i}{du^i}\varphi^{-1}(u) \geq 0, \text{ for } i = 1, \ldots, d - 2$$

Also $\varphi$ is generator for $i$-dimensional copulas $C^{(i)}$ of dimensions $i = 2, \ldots, d - 1$. Some of the most well-known families and their corresponding generators are displayed in Table (1). Archimedean copulas have some algebraic properties that we will use for this research.

Proposition 1. Let $C^{(d)}$ an Archimedean copula with generator $\varphi$. Then

1. $C^{(d)}$ is symmetric, i.e. $C(u_1, \ldots, u_d) = C(u_{\sigma(1)}, \ldots, u_{\sigma(d)})$ for all $u \in [0,1]^d$ and all permutations $\sigma(1), \ldots, \sigma(d)$ of $1, \ldots, d$;

2. $C^{(d)}$ is associative, i.e.

$$C(C(u_1, u_2), u_3) = C(u_1, C(u_2, u_3))$$

for all $u_1, u_2, u_3 \in [0,1]$ and similarly for higher dimensions.

3. Let $c > 0$ be a real constant; then $c\varphi$ is also a generator of $C^{(d)}$

Note that, the density of the copula $C^{(d)}$, namely $c(\cdot)$, is given by

$$c(u_1, \ldots, u_d) = \varphi^{-1(d)}\big(\varphi(u_1) + \cdots + \varphi(u_d)\big) \prod_{i=1}^{d} \varphi'(u_i) \tag{4}$$

Another important concept for copulas is the Kendall distribution function, that can be seen as the distribution of the level curves (surfaces). If ($U_1, \ldots, U_d$) is a random vector of uniform variables with distribution function given by a copula $C^{(d)}$, then the distribution function of the variable $C(U_1, \ldots, U_d)$ is known as the Kendall distribution function, specifically:

$$K(t) = P(C(U_1, \ldots, U_d) \le t)$$

In the case of Archimedean copulas, the Kendall distribution for the bivariate case is $K(t) = t - \varphi(t)/\varphi'(t)$ (Genest & Rivest, 1993). For higher dimensions, Barbe, Genest, Ghoudi, and Rémillard (1996) presented the terms for Kendall process for an Archimedean copula $C^{(d)}$, with generator $\varphi$ recursively by:

$$K(t) = t + \sum_{i=1}^{d-1} (-1)^i \frac{\varphi(t)^i}{i!} f_{i-1}(t) \tag{5}$$

where the auxiliary functions $f_i(t)$ are defined recursively for $i = 1, \ldots, d-2$:

$$f_0(t) = \frac{1}{\varphi'(t)} \tag{6}$$

$$f_i(t) = \frac{f'_{i-1}(t)}{\varphi'(t)} \tag{7}$$

Observe that;

$$f_i(t) = \frac{d^{i+1}}{ds^{i+1}} \varphi^{-1}(s) \bigg|_{s=\varphi(t)} \tag{8}$$

One can show that the density for Kendall distribution is given by:

$$k(t) = (-1)^{d-1} \frac{\varphi(t)^{d-1}}{(d-1)!} \varphi'(t) f_{d-1}(t) \tag{9}$$

Existence for this distribution is guaranteed by the regularity conditions. For general properties about Kendall processes, refer to Barbe et al. (1996); Imlahi and Chakak (1999); Nelsen, Quesada-Molina, Rodríguez-Lallena, and Ubeda-Flores (2003) and references therein.

When studying the dependence structure between two or more variables, one can present measures of association like the rank-correlation coefficient Kendall's tau which measures the concordance. In terms of copulas, the Kendall's tau for a pair of standard uniform variables, $U$ and $V$ with copula $C^{(2)}$ is given by:

$$\tau_{U,V} = 4E[C(U,V)] - 1$$

Moreover, if the copula is Archimedean, the Kendall's tau can be expressed in terms of its dependence parameter $\theta$ as (Joe, 1997):

$$\tau_{U,V} = \tau_\theta = 4 \int_0^1 \frac{\varphi(t)}{\varphi'(t)} dt + 1$$

| Family | $\tau$ | $\lambda_l$ | $\lambda_u$ |
|---|---|---|---|
| Clayton | $\theta/(\theta + 2)$ | $2^{-1/\theta}$ | 0 |
| Frank | $1 + 4(D_1(\theta) - 1)/\theta$[1] | 0 | 0 |
| Gumbel | $(\theta - 1)/\theta$ | 0 | $2 - 2^{1/\theta}$ |
| Joe | $1 - 4\sum_{k=1}^{\infty} \frac{1}{k(\theta k + 2)(\theta(k-1) + 2)}$ | 0 | $2 - 2^{1/\theta}$ |
| Ali-Mikhail-Haq | $1 - \frac{2(\theta + (1-\theta)^2 \log(1-\theta))}{3\theta^2}$ | 0 | 0 |

Table 2 Kendall's tau and tail-dependence coefficients for some Archimedean Copulas.

[1]$D_1$ refers to the Debye function of order one: $D_1(\theta) = \frac{1}{\theta} \int_0^\theta \frac{t}{e^t - 1} dt$.

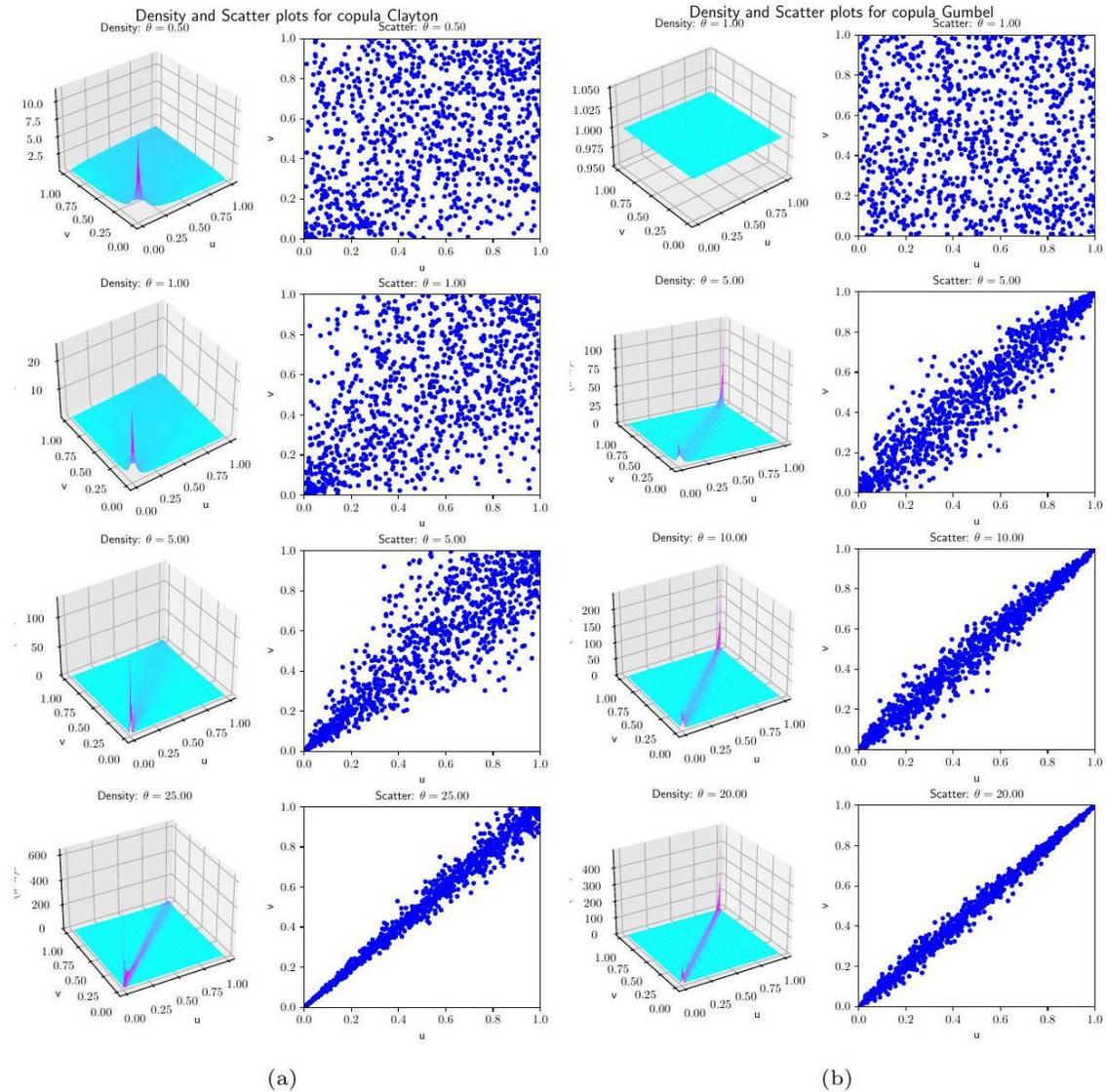

Fig. 1 Density and scatter plots for 2-dimensional Clayton and Gumbel families with different dependence parameters.

Note that these coefficients vary in $[-1,1]$ meaning a strong relationship for both variables in the sense that large values of one are related to large values of the left one. The concept of tail-dependence coefficients is recalled below. This coefficient measures the probability that one variable takes values in its tail, given that the left one also takes values in its tail. That is, it quantifies the likelihood of joint extreme events between the two variables. Formally, if $F_1$ and $F_2$ are the distribution functions of the continuous random vector (

$X_1, X_2$ ) with copula $C^{(2)}$, then the lower and upper tail-dependence coefficient are given by:

$$\lambda_l = \lim_{t \to 0^+} P(X_2 \leq F_2^{-1}(t) \mid X_1 \leq F_1^{-1}(t)) = \lim_{u \to 0^+} \frac{C(u,u)}{u},$$

$$\lambda_u = \lim_{t \to 1^-} P(X_2 > F_2^{-1}(t) \mid X_1 > F_1^{-1}(t)) = \lim_{u \to 1^-} \frac{\hat{C}(u,u)}{u},$$

where $\hat{C}$ is the survival copula of $C^{(2)}$ given by $\hat{C}(u,v) = u + v - 1 + C(1-u, 1-v)$ (McNeil et al., 2015). For bivariate Archimedean copulas with generator $\varphi$, the corresponding tail-dependence coefficients are given by:

$$\lambda_l = \lim_{t \to \infty} \frac{\varphi^{-1}(2t)}{\varphi^{-1}(t)} = 2 \lim_{t \to \infty} \frac{\varphi^{(-1)'}(2t)}{\varphi^{(-1)'}(t)}$$

$$\lambda_u = 2 - \lim_{t \to 0^-} \frac{1 - \varphi^{-1}(2t)}{1 - \varphi^{-1}(t)} = 2 - 2 \lim_{t \to 0^-} \frac{\varphi^{(-1)'}(2t)}{\varphi^{(-1)'}(t)}$$

The parameters for some Archimedean generators can be found in Table 2.

# 3 Results

Recently, a definition of Copula-based Value at Risk (CCVaR) was presented in Molina Barreto and Ishimura (2023). However, results for Archimedean copulas were restricted to the bivariate case. An extension of this definition to the multivariate context is proposed, and some of its properties are investigated. Let $\boldsymbol{X} = (X_1, \ldots, X_d)^t$ a random vector of dimension $d (d \geq 2)$ with joint distribution function $H(\boldsymbol{X}) = P(X_1 \leq x_1, \ldots, X_d \leq x_d)$ and its respective marginal distributions $F_{X_i}(x) = P(X_i \leq x)$ for $i = 1, \ldots, d$. It can be considered that each component represents the loss for each risk factor in a portfolio. Let $\boldsymbol{\lambda} = (\lambda_1, \ldots, \lambda_d)^t$ be a vector representing the weights for the portfolio, where $\sum_{i=1}^{d} \lambda_i = 1$ and $0 \leq \lambda_i < 1$ for $i = 1, \ldots, d$. To avoid unfavorable situations, it is assumed that the regularity conditions are satisfied (see Cousin & Di Bernardino, 2014, Assumption 2.1). Because of Sklar's theorem, $H$ can be represented by a copula $C^{(d)}$ and for $0 \leq \beta < 1$, setting:

$$\mathcal{U}_\beta^{(d)} := \{(u_1, \ldots, u_d) \in [0,1]^d \mid C^{(d)}(u_1, \ldots, u_d) \geq \beta\}. \tag{10}$$

Considering the above set, as the unfavorable set as in the definition of Prékopa (2012), a Copula-based Conditional Value at Risk for a confidence level $\beta$ can be defined as stated in the following definition.

Definition 4. Let $\boldsymbol{X} = (X_1, \ldots, X_d)^t$ be a continuous random vector of dimension $d(d \geq 2)$, $C^{(d)}$ a copula for $\boldsymbol{X}$ and $\boldsymbol{\lambda} = (\lambda_1, \ldots, \lambda_d)^t$ a vector representing the weights for the portfolio, where $\sum_{i=1}^{d} \lambda_i = 1$ and $0 \leq \lambda_i < 1$. A Copula-based Conditional Value at Risk (CCVaR) of $X$ at level $\beta (0 \leq \beta < 1)$ is given by:

$$\text{CCVaR}_\beta(\boldsymbol{X}) = \frac{\int \cdots \int_{\mathcal{U}_\beta^{(d)}} \left(\lambda_1 F_{X_1}^{(-1)}(u_1) + \cdots + \lambda_d F_{X_d}^{(-1)}(u_d)\right) dC(u_1, \ldots, u_d)}{\int \cdots \int_{\mathcal{U}_\beta^{(d)}} dC(u_1, \ldots, u_d)} \tag{11}$$

where $\mathcal{U}_\beta^{(d)}$ is given by (10).

It is important to note that for a general copula, the numerator may be infinite or denominator zero. However, under the regularity conditions, it can be safely assumed that the integral is well-defined and that the denominator is non-zero. It is noted that the preceding definition is based on the following observation:

$$\text{CCVaR}_\beta(\boldsymbol{X}) = E^C\left[\boldsymbol{\lambda}^t \boldsymbol{F}_X^{(-1)} \mid \mathcal{U}_\beta^{(d)}\right] \tag{12}$$

Here, $E^C\left[\cdot \mid \mathcal{U}_\beta^{(d)}\right]$ denotes the expected value w.r.t copula $C^{(d)}$. The previous definition can be seen as an application of copula at the MCVaR by Lee and Prékopa (2013). This definition may be interpreted as a linear combination of the terms for the lower-orthant Conditional Tail Expectation (CTE), implying that both measures share key properties (see Cousin & Di Bernardino, 2014, Eq. 8). However, the resulting measure is vector-valued, in contrast to the scalar CCVaR. Such a vector-valued formulation can be advantageous, as it represents the total loss when all unfavorable events have occurred. The vector $\boldsymbol{\lambda}$, not only represents the portfolio weights but also multiplies each loss, thereby enabling the aggregation of different asset losses on a common scale.

Proposition 2. A copula-based conditional Value at Risk CCVaR defined by (11) verifies the following properties.

i. $\text{CCVaR}_\beta(\mathbf{0}) = 0$,
ii. $\text{CCVaR}_\beta(\mathbf{X} + k\mathbf{e}) = \text{CCVaR}_\beta(\mathbf{X}) + k$ ($k \in \mathbb{R}, \mathbf{e} = (1, \ldots, 1)$),
iii. If every component in $\mathbf{X}_1$ is smaller than the corresponding component in $\mathbf{X}_2$ in stochastic order, then we have
$$\text{CCVaR}_\beta(\mathbf{X}_1) \leq \text{CCVaR}_\beta(\mathbf{X}_2).$$
iv. $\text{CCVaR}_\beta(s\mathbf{X}) = s\text{CCVaR}_\beta(\mathbf{X})$ ($s > 0$).
v. If $\mathbf{X}_1, \mathbf{X}_2$ are continuous and all components in $\mathbf{X}_1, \mathbf{X}_2$ are independent, then we have
$$\text{CCVaR}_\beta(\mathbf{X}_1 + \mathbf{X}_2) \leq \text{CCVaR}_\beta(\mathbf{X}_1) + \text{CCVaR}_\beta(\mathbf{X}_2).$$

The property (v) known as sub-additivity, is not always true. For a counterexample in the general case, refer to Theorem 8 in Lee and Prékopa (2013).

Specifically, they provided a counterexample when $X_1$ and $X_2$ are discrete with certain distributions. According to them, sub-additivity can be seen as both good and bad *M&A* cases, since not all these deals will be successful. For negative cases, the risk will not be reduced, which means that MCVaR will not be sub-additive. So, $\text{CCVaR}_\beta(X)$ is a coherent risk measure if satisfies the hypothesis in the Prop. 2.

## 3.1 CCVaR for Archimedean copulas

Now the main results of this research are stated. In the case of Archimedean copula, the CCVaR can expressed in terms of the higher order derivative of the inverse generator.

Theorem 3. Let $X = (X_1, \ldots, X_d)$ be a non-negative random vector ($d \geq 2$), with joint distribution function is provided by an Archimedean copula $C^{(d)}$ as defined in (3), where the generator $\varphi$ is $d$-monotonic. Then, for a confidence level $\beta$, the $\text{CCVaR}_\beta(X)$ in (11) can be expressed as:

$$\text{CCVaR}_\beta(X) = \frac{\int_\beta^1 \left(\lambda_1 F_{X_1}^{(-1)}(t) + \cdots + \lambda_d F_{X_d}^{(-1)}(t)\right)\varphi'(t)h_{d-1}(t,\beta)dt}{1 - K_\varphi^{(d)}(\beta)} \tag{13}$$

where $K_\varphi^{(d)}(\cdot)$ is the Kendall distribution function for the copula $C^{(d)}$, and $h_{d-1}(\cdot,\cdot)$ is defined by:

$$h_{d-1}(t,\beta) = f_0(t) - f_0(\beta) - \sum_{i=1}^{d-2} \frac{f_i(\beta)}{i!}[\varphi(t) - \varphi(\beta)]^i \tag{14}$$

and $f_i(t)$ are defined as (6) and (7)

The proof can be found in Appendix A.

The previous theorem is a generalization to several variables of the one presented in Molina Barreto and Ishimura (2023) (see Theorem 3). In this case for $d = 2$, $h_1(t,\beta) = f_0(t) = 1/\varphi'(t)$ and $K_\varphi^{(2)}(t) = t - \varphi(t)/\varphi'(t)$, then applying the Theorem (3) the result is immediate.

To make direct use of the previous theorem, an efficient way to compute all derivatives of the inverse of the generating function up to order $d - 1$ is required. Fortunately, explicit forms for these derivatives for all the families considered in this study can be found in Hofert, Mächler, and McNeil (2012). See Table 3 for a list of the explicit forms of $f_i(t)$.

Remark 1. If the independence copula is considered, the CCVaR is reduced to the MCVaR by Lee and Prékopa (2013). If the generator is $\varphi(t) = -\log t$, the resulting copula is the product copula:

$$C(u_1, \ldots, u_d) = \prod_{i=1}^{d} u_i$$

that represents the independence structure between variables $X_1, \ldots, X_d$. The auxiliary functions are $f_i(t) = (-1)^{i+1} t$, so the terms $h_{d-1}(t, \beta)$ reduce to:

$$h_{d-1}(t, \beta) = -t + \beta + \beta \sum_{i=1}^{d-2} \frac{(-1)^i}{i!} (-\log t + \log \beta)^i$$

Here, $K^{(d-1)}(t)$ represents the Kendall distribution function for the independence copula of dimension $d-1$. Then the MCVaR$_\beta$ can be written as:

$$\text{MCVaR}_\beta(X) = \frac{1}{1 - K^{(d)}(\beta)} \int_\beta^1 \left( \sum_{i=1}^{d} \lambda_i F_{X_i}^{(-1)}(t) \right) \left[ 1 - K^{(d-1)}\left(\frac{\beta}{t}\right) \right] dt \qquad (15)$$

Example 1. Clayton family of copulas are widely used to modeling the lower tail dependence. Consider the generator for the Clayton copula $\varphi(t) = t^{-\theta} - 1$ and its derivatives of its inverse of order $d$, $f_i(t) = (-1)^{i+1} \frac{\Gamma(i+1+1/\theta)}{\Gamma(1/\theta)} t^{\theta(i+1)+1}$ for $i = 1, \ldots, d-2$, and $\Gamma(\cdot)$ is the gamma function. The corresponding terms $h_{d-1}(t, \beta)$ are:

$$h_{d-1}(t, \beta) = \frac{1}{\theta}(-t^{\theta+1} + \beta^{\theta+1}) - \sum_{i=1}^{d-2} \frac{(-1)^{i+1} \Gamma(i+1+1/\theta)}{i! \Gamma(1/\theta)} \beta^{\theta(i+1)+1} [t^{-\theta} - \beta^{-\theta}]$$

Then the corresponding CCVaR$_\beta(X)$ is reduced to:

$$\text{CCVaR}_\beta(X) = \frac{1}{1 - \beta - \sum_{i=1}^{d-1} \frac{(-1)^i \Gamma(i+1/\theta)}{(i-1)! \Gamma(1/\theta)} \beta^{\theta i+1}} \int_\beta^1 \left( \sum_{i=1}^{d} \lambda_i F_{X_i}^{(-1)}(t) \right)$$

$$\left( 1 - \left(\frac{\beta}{t}\right)^{\theta+1} + \theta \left(\frac{\beta}{t}\right)^{\theta+1} \sum_{i=1}^{d-2} \frac{(-1)^{i+1} \Gamma(i+1+1/\theta)}{i! \Gamma(1/\theta)} \left(\frac{\beta}{t}\right)^{\theta i} \left[ \left(\frac{\beta}{t}\right)^\theta - 1 \right]^i \right) dt$$

As a particular property of the Clayton family, is that if the parameter $\theta$ tends to infinity, the dependence structure converges to the Comonotonic copula, that is, the perfectly increasing dependence. Knowing that $(\beta/t) < 1$, the limit inside the integral exists and the Kendall's distribution function tends to the identity in $[0,1]$. Then $\text{CCVaR}_\beta$ results in:

$$\text{CCVaR}_\beta(X) = \frac{\int_\beta^1 \left(\sum_{i=1}^d \lambda_i F_{X_i}^{(-1)}(t)\right)dt}{1-\beta} \tag{16}$$

i.e., $\text{CCVaR}_\beta$ collapses to the weighted sum of the corresponding univariate $\text{CVaR}_\beta$ for each component of the vector $X$.

Table 3 Auxiliary functions $f_i(t)$ giving the $(i+1)$-th derivative of the inverse generator $\varphi^{-1}(t)$ for common Archimedean copula families.

| Family | $(-1)^{i+1} f_i(t)$ |
|---|---|
| Independence | $t$ |
| Clayton | $\dfrac{\Gamma(i+1+1/\theta)}{\Gamma(1/\theta)} t^{\theta(i+1)+1}$ |
| Frank [1] | $\dfrac{1}{\theta} \text{Li}_{-i}(1-e^{-\theta t})$ |
| Gumbel [2] | $\dfrac{t}{(-\log t)^{\theta(i+1)}} P_{i+1,\theta}^G\left((-\log t)^{\alpha\theta}\right)$ |
| Joe [3] | $\dfrac{1-(1-t)^\theta}{\theta(1-t)^{\theta-1}} P_{i+1,\theta}^J\left(\dfrac{1-(1-t)^\theta}{(1-t)^\theta}\right)$ |
| Ali-Mikhail-Haq [1] | $\dfrac{1-\theta}{\theta} \text{Li}_{-(i+1)}\left(\dfrac{\theta}{t}(\theta(1-t)-1)\right)$ |

[1] $\text{Li}_s(z) = \sum_{k=1}^\infty z^k/k^s$ is the polylogarithm of order $s$ at $z$.

[2] Here $P_{i+1,\theta}^G(x)$ is a polynomial of degree $i+1$ where

$P_{i,\theta}^G(x) = \sum_{k=1}^d a_{ik}^G(\theta) x^k$

$a_{ik}^G(\theta) = (-1)^{i-k} \sum_{j=k}^i \theta^{-j} s(i,j) S(j,k) = \frac{i!}{k!} \sum_{j=1}^k \binom{k}{j}\binom{\alpha j}{i}(-1)^{i-j}, k \in \{1, \dots, i\}$, and $s$ and $S$ denote the Stirling numbers of the first kind and the second kind, respectively.

[3] Here $P_{i+1,\theta}^J(x)$ is a polynomial of degree $i$ where

$P_{i,\theta}^J(x) = \sum_{k=1}^{i} a_{ik}^J(\theta) x^{k-1}$

$a_{ik}^J(\theta) = S(i,k)(k-1-1/\theta)_{k-1} = S(i,k)\frac{\Gamma(k-\alpha)}{\Gamma(1-\alpha)}, k \in \{1,\ldots,i\}$ respectively.

Notice that the standard Conditional Value at Risk can be interpreted as the mean loss given that the loss variable is comonotonic with the risk factor. Thus, for a single asset with a comonotonic copula, $\text{CCVaR}_\beta$ coincides with the standard univariate CVaR. However, for a bivariate Clayton copula, an increase in the dependence parameter results in a decrease in $\text{CCVaR}_\beta$. In contrast to the standard CVaR, where comonotonicity yields the worst risk measure, $\text{CCVaR}_\beta$ is minimized under comonotonicity. This somewhat surprising result is consistent with the findings in Cousin and Di Bernardino (2014, see Section 2.2). This behavior can be observed in Figure 3 for higher dimensions in the Clayton and Gumbel families. In essence, as the parameter $\theta$ increases, the integration region $\mathcal{U}_\beta$ expands (eventually approaching a rectangle with upper vertex at $(1,\ldots,1)$, yet the density concentrates along the diagonal of the hypercube $[0,1]^d$, so that the effective integration domain becomes larger while the denominator, the term $1 - K(t)$, as function of $\theta$, increases faster than the numerator for the previous mentioned copulas.

It is expected that a risk measure will be a non-decreasing function of the confidence level $\beta$ as it is $\text{VaR}_\beta$ and $\text{CVaR}_\beta$ for continuous distributions. The following proposition states that the same property is shared with the CCVaR for Archimedean copulas.

Proposition 4. Let $\boldsymbol{X} = (X_1,\ldots,X_d)$ be a random vector with distribution given by a copula $C^{(d)}$. If it is Archimedean, then $\text{CCVaR}_\beta(\boldsymbol{X})$ is a non-decreasing function of the risk level $\beta$. This is, if $0 < \beta_1 \leq \beta_2 < 1$, then

$$\text{CCVaR}_{\beta_1}(\boldsymbol{X}) \leq \text{CCVaR}_{\beta_2}(\boldsymbol{X}) \qquad (17)$$

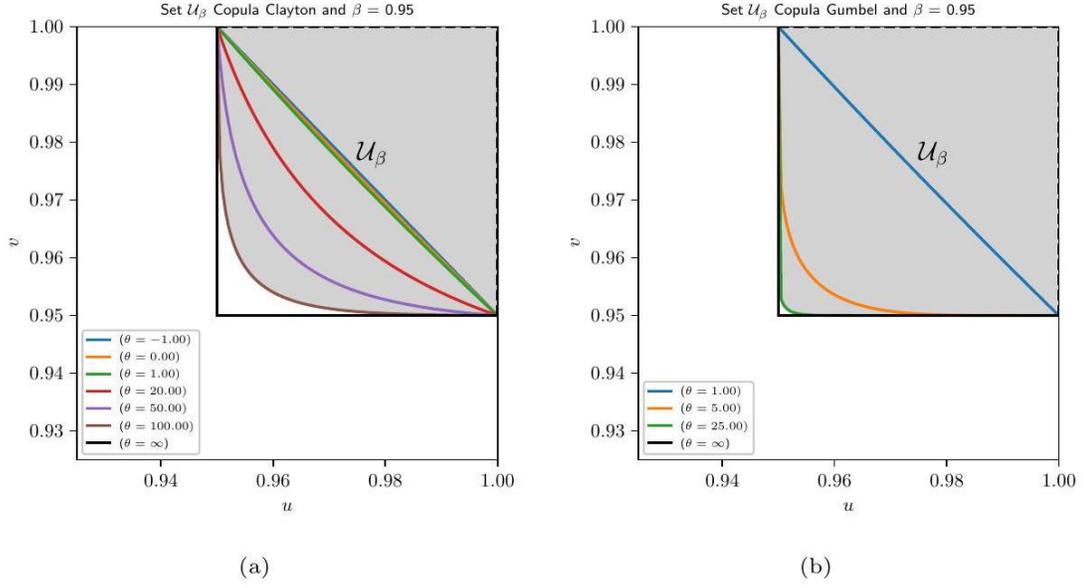

Fig. 2 Sets $\mathcal{U}_\beta$ for Clayton and Gumbel copulas with several dependence parameters $\theta$.

In the context of risk management, safety loading may refer to the extra buffer or margin that helps protect against the likelihood for severe losses in the tail of the loss distribution being useful as a risk cushion beyond the average or threshold-based estimates. In this sense, the CCVaR satisfies this property.

Proposition 5. Let $X$ be a d-dimensional random vector with Archimedean copula $C^{(d)}$, and $0 \leq \lambda_i \leq 1$ such that, $\sum_{i=1}^d \lambda_i = 1$, then the Safety Loading property is satisfied. This is, for $0 < \beta \leq 1$,

$$\text{CCVaR}_\beta(X) \geq \sum_{i=1}^d \lambda_i E[X_i] \qquad (18)$$

For instance, for standard uniform margins and equal weights, using the copulas listed in Table 1, one can observe in Figure 3 that $\text{CCVaR}_\beta(X) \geq 0.5$, in agreement with the safety loading property.

One of the fundamental relationships between the univariate VaR and CVaR is that for any risk level $\beta$, $\text{VaR}_\beta(X) \leq \text{CVaR}_\beta(X)$. For the multivariate extension, a similar property holds provided that the joint distribution of $X$ is quasi-concave. A function $F$ is quasi-concave if for any $x, y$, and for all $p \in (0,1)$,

$$F(px + (1-p)y) \geq \min\{F(x), F(y)\}$$

Fortuitously, for Archimedean copulas, the upper-level sets are convex, ensuring quasiconcavity. By virtue of Proposition 2.6 in Cousin and Di Bernardino (2014), the following proposition is stated.

Proposition 6. If the d-dimensional random vector $X$ with an Archimedean copula $C^{(d)}$, $\lambda = (\lambda_1, \ldots, \lambda_d)^t$ a vector representing the weights for the portfolio, where $\sum_{i=1}^{d} \lambda_i = 1$ and $0 \leq \lambda_i < 1$, and $\text{VaR}_\beta(X_i)$ satisfies the sub-additivity property for $i = 1, \ldots, d$, then

$$\text{VaR}_\beta(\lambda^t X) \leq \text{CCVaR}_\beta(X) \qquad (19)$$

Proof of this proposition is trivial if we consider that $\text{CCVaR}_\beta$ can be written as a linear combination of the components of the Conditional Tail-Expectation by Cousin and Di Bernardino (2014). See also, Hürlimann (2014).

The behavior of $\text{CCVaR}_\beta$ with respect to the copula parameter $\theta$, which represents the level of dependence between the variables, depends strongly on the specific copula family. For example, for $d = 2$, (Cousin & Di Bernardino, 2014, see Corollary 2.2) it was mentioned that for Clayton, Frank, Gumbel, and Ali-Mikhail-Haq copulas, an increase in $\theta$ yields a decrease in the lower-orthant CTE components. However, our empirical examination of the Ali-Mikhail-Haq copula with uniform margins and $\beta = 0.95$ suggests that this holds only for $\theta \in (0,1)$, not for all admissible $\theta$. A similar observation holds for the Gumbel copula, where the CTE is monotonic only for certain values of $\theta \gg 1$, as shown in Figure 4. These findings are summarized in the next proposition.

Proposition 7. Let $X$ be a bidimensional random vector that satisfies the regularity conditions, with marginal distribution $F_i$ for $i = 1,2$. If the corresponding copula $C^{(2)}$ is of one the following Archimedian families:

- Clayton with $\theta \in (-1, \infty)$
- Frank with $\theta \in (0, \infty)$
- Gumbel with $\theta \in (2, \infty)$
- AMH with $\theta \in (0,1)$

then $\text{CCVaR}_\beta(X)$ is a decreasing function of $\theta$.

Remark 2. According to the above proposition, relationships between the CCVaR and MCVaR can be clearly stated, as the independence copula can be treated as a special or limiting case of some of the aforementioned copulas. For instance, consider Gumbel ( $\theta = 1$ ), and AMH ( $\theta = 0$ ) copulas, for any $\theta^* \geq \theta$, in the interval shown in prop. (7), then the corresponding $\text{CCVaR}_\beta$ satisfies:

$$\text{CCVaR}_\beta(X) \leq \text{MCVaR}_\beta(X) \tag{20}$$

that corresponds to the Examples 1&3 in Molina Barreto and Ishimura (2023). However, a general assertion for any arbitrary dimension does not appear to hold for certain copula families. For example, as shown in Figure 3, for Joe and Frank copulas there is no clear monotonic relationship between $\text{CCVaR}_\beta$ and $\theta$ for all $t \in (\beta, 1)$.

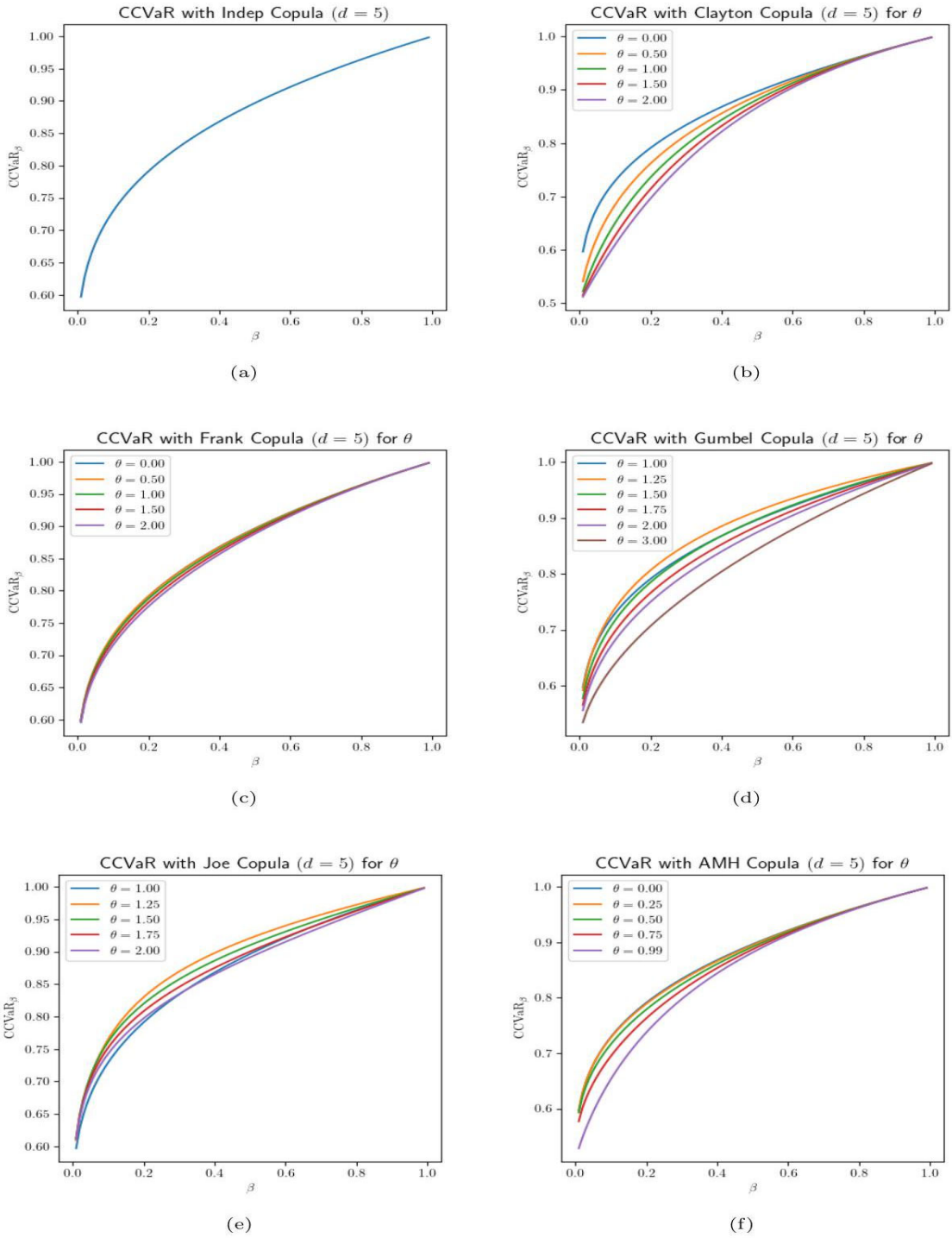

Fig. 3 CCVaR$_\beta$ with respect $\beta$ for some dependence levels $\theta$, $d=5$ and standard uniform margins with equal weights. (a) Independence (b) Clayton (c) Frank (d) Gumbel (e) Joe (f) AMH.

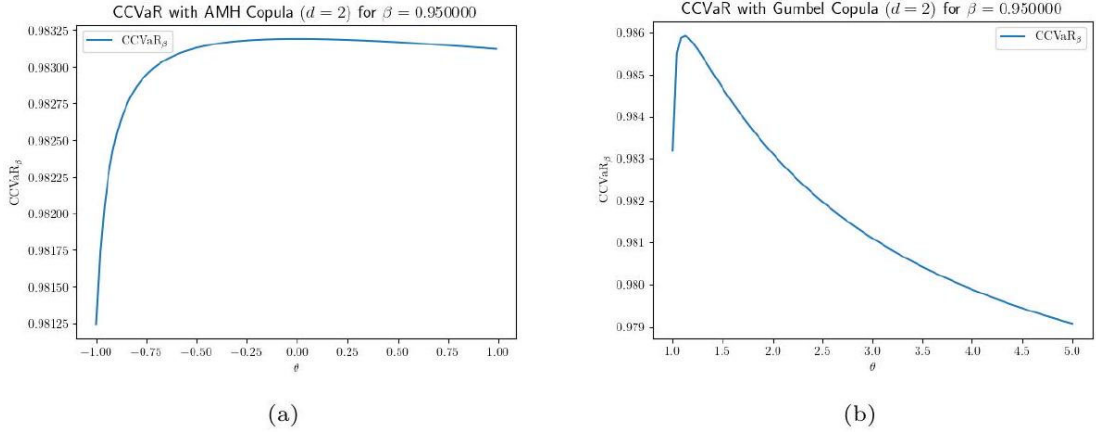

Fig. 4 CCVaR$_\beta$ with respect dependence level $\theta$, $\beta = 95\%$, $d = 2$ and standard uniform margins with $\lambda_1 = 1$. (left) AMH (right) Gumbel.

## 3.2 Numerical studies

In this section, numerical results applying the calculations of the copula based Conditional Value at Risk are presented and compared with other classical approaches for estimation of Value at Risk and Conditional Value at Risk via copulas using real data from the market. Part of the estimation was performed in $R$, and most of the implementation was written in Python. Codes can be downloaded from the author's Github repository.

## 3.2.1 Descriptive Statistics and Margin modeling

Random vectors from Archimedean copulas are equi-correlated, so considering widely diversified portfolios is not adequate in this setting (Savu & Trede, 2008). For this study, a set of 7 stocks listed in the Nikkei 225 index corresponding to the machinery sector were studied. The stocks were chosen because they belong to the same sector, which allows for a more straightforward application of copula models. The companies and their symbols are listed in table 4. Daily closing prices from Dec. 19, 2018, to Jul. 18, 2024, are taken for a total of $T = 1360$ observations. This period notably includes the COVID-19 pandemic, which provides a good dataset for this analysis. Data was obtained from Yahoo! Finance website. Daily negative log-returns are computed, and descriptive statistics can be found in table 5. Some of these series exhibit excess of kurtosis indicating heavy tails for extreme losses, so considering heavy-tailed distributions would be necessary.

Models AR(1)-GARCH(1,1) for each series were fitted considering different distribution for the standardized residuals, such as Standard Normal, Student-T, and Skewed-T. This choice was made primarily due to the simplicity and ease of implementation of this model, which is known to handle volatility clustering effectively. Although various ARMA-GARCH specifications were explored, the results did not vary significantly, so AR(1)-GARCH(1,1) was chosen to maintain consistency across the margins. These results are reported in Appendix B .

Table 4 List of symbols of Machinery companies belonging to Nikkei 225

| Symbol | Company Name |
| --- | --- |
| 6103.$T$ | OKUMA CORP. |
| 6113.$T$ | AMADA CO., LTD. |
| 6273.$T$ | SMC CORP. |
| 6301.$T$ | KOMATSU LTD. |
| 6302.$T$ | SUMITOMO HEAVY IND., LTD. |
| 6305.$T$ | HITACHI CONST. MACH. CO., LTD. |
| 6326.$T$ | KUBOTA CORP. |

Table 5 Descriptive statistics for negative log-returns of assets in G1

|  | 6103.T | 6113.T | 6273.T | 6301.T | 6302.T | 6305.T | 6326.T |
| --- | --- | --- | --- | --- | --- | --- | --- |
| Mean | -0.0421 | -0.0579 | -0.0626 | -0.0644 | -0.026 | -0.0532 | -0.0285 |
| Std | 2.068 | 1.8095 | 1.9765 | 1.889 | 1.928 | 2.2064 | 1.8512 |

| Min | -13.7904 | -8.9907 | -8.8831 | -10.8989 | -11.6993 | -14.9036 | -7.0551 |
| --- | --- | --- | --- | --- | --- | --- | --- |
| Median | 0 | -0.0836 | -0.0471 | -0.096 | -0.0691 | -0.1358 | -0.1188 |
| Max | 9.6372 | 7.3277 | 9.3228 | 10.6062 | 8.1056 | 18.6178 | 13.4865 |
| Kurtosis | 4.0149 | 1.6061 | 1.5375 | 3.5864 | 2.9894 | 9.1666 | 3.2724 |
| Skewness | -0.3702 | -0.0692 | -0.0326 | -0.0162 | -0.1551 | 0.7898 | 0.543 |

### 3.2.2 Copula estimation

Estimation of the parameters of the copula and their corresponding margins can be done in two steps with the method known as inference functions for margins (IFM). Also, the method of pseudo maximum likelihood proposed in Genest, Ghoudi, and Rivest (1995), can be used in the case the marginal distributions functions are approximated by their empirical distribution. For simplicity, it is assumed that the realizations of the vector are independent, in which case the IFM method can be applied. Specifically, estimation for Archimedean copulas listed here, can be computed effectively thanks to the closed-form density functions given in Hofert, Mächler, and McNeil (2011). If $\boldsymbol{u}_i$, $i = 1, \ldots, T$ are realizations of a random sample $\boldsymbol{U}_{ij}, i = 1, \ldots, T$, and $j = 1, \ldots, d$, the log-likelihood function is defined by

$$\ln L(\theta; U_{ij}) = \sum_{i=1}^{T} \ln c(U_{i1}, \ldots, U_{id}) \qquad (21)$$

where $c(\cdot)$ represents the density of the Archimedean copula. The maximum likelihood estimator of $\theta$ is

$$\hat{\theta} = \operatorname{argmax} \ln (\theta; U_{ij}) \qquad (22)$$

whose estimation is performed by numerical routines.

In table 6, the results for the estimation of the copula parameter are presented for each type of margins with their corresponding standard errors and the negative value of the log-likelihood. The distance $d_\gamma$ between the estimated and the empirical copula in the upper tail $[\gamma, 1]^d$ was also estimated ($\gamma = 0.8, 0.9$). Lesser values indicate better fits in the case of extreme events. Also, goodness-of-fit tests for the copula based on the ranks were performed. In this case, $S_n$ represents the Cramér-von Mises statistic. However, as Savu and Trede (2008) remarked, this type of tests lose power as the dimension of the copula increases. Then, calculations for their statistic $T_n$ and p-values were also implemented, which are based on the classical $\chi^2$ projecting the multivariate data into the unit interval. In this regard, the best estimates are given by the Gumbel copula with skewed-t innovations. It is also observed that this estimate gives the minimum distances to the empirical copula in the upper tail. The second best results are from Joe copula.

| Copula | Margins | $\hat{\theta}$ | Std. Error | Log-lik. | $d_{0.8}$ | $d_{0.9}$ | $S_n$ | $T_n$ | p-value |
|---|---|---|---|---|---|---|---|---|---|
| Clayton | Normal | 0.7435 | 0.0153 | -2066.2270 | 0.0145 | 0.0059 | 6.7669 | 1.779E + 07 | <0.001 |
| | Std-t | 0.7432 | 0.0153 | -2066.0628 | 0.0107 | 0.0035 | 6.8003 | 1.785E + 07 | <0.001 |
| | Ske-t | 0.7429 | 0.0153 | -2065.2255 | 0.0124 | 0.0051 | 6.8177 | 1.235E + 08 | <0.001 |
| Frank | Normal | 3.9468 | 0.0682 | -2475.9453 | 0.0050 | 0.0024 | 1.7042 | 1.406E + 04 | <0.001 |
| | Std-t | 3.9489 | 0.0681 | -2481.9624 | 0.0024 | 0.0012 | 1.7096 | 8.466E + 04 | <0.001 |
| | Ske-t | 3.9484 | 0.0681 | -2482.2979 | 0.0032 | 0.0020 | 1.7151 | 8.472E + 04 | <0.001 |
| Gumbel | Normal | 1.5701 | 0.0129 | -2491.1205 | 0.0010 | 0.0004 | 2.7614 | 1.333E + 02 | 0.007 |
| | Std-t | 1.5714 | 0.0129 | -2502.8112 | 0.0003 | 0.0002 | 2.7464 | 1.243E + 02 | 0.03 |
| | Ske-t | 1.5715 | 0.0129 | -2503.8828 | 0.0003 | 0.0001 | 2.7491 | 1.217E + 02 | 0.05 |
| Joe | Normal | 1.8092 | 0.0199 | -2018.6558 | 0.0005 | 0.0004 | 5.4161 | 2.488E + 02 | 0.001 |

|  |  |  |  |  |  |  |  |  |
|---|---|---|---|---|---|---|---|---|
|  | Std-t | 1.8129 | 0.0199 | -2033.2794 | 0.0004 | 0.0005 | 5.3801 | 2.527E + 02 | <0.001 |
|  | Ske-t | 1.8134 | 0.0199 | -2035.0799 | 0.0003 | 0.0003 | 5.3798 | 2.421E + 02 | <0.001 |
| AMH | Normal | 0.9304 | 0.0053 | -2282.1013 | 0.0129 | 0.0054 | 5.4796 | 6.458E + 06 | <0.001 |
|  | Std-t | 0.9302 | 0.0053 | -2283.2524 | 0.0092 | 0.0032 | 5.5126 | 6.475E + 06 | <0.001 |
|  | Ske-t | 0.9301 | 0.0053 | -2282.6991 | 0.0108 | 0.0047 | 5.5270 | 4.438E + 07 | <0.001 |

Table 6 Results for Copula Models and Marginal Distributions

### 3.2.3 VaR, CVaR and CCVaR estimation

After the estimates for the margins and the copula have been obtained, CCVaR estimation for the Archimedean copulas in Table 1 is carried out via numerical integration of equation (13) with risk levels $\beta = 0.95 \& 0.99$. For simplicity, all weights $\lambda_i$ are considered equal. As is observed in the results, the values of CCVaR show little variability with respect to copula choice. However, there were certain scenarios where numerical instability became evident; for example, for the AMH copula at the most extreme values such as $\beta = 0.99$. In this case, it is recalled that the estimation of the $CCVaR_\beta$ depends on the function

$h_{d-1}(t, \beta)$ which is expressed in terms of the auxiliary functions $f_i(t)$. Should issues arise in the evaluation of this function for extreme values, the approach proposed by Hofert, Mächler, and McNeil (2013, Sec. 3.5) is applied. Accordingly, a Monte Carlo approximation with m realizations is given by:

$$f_i(t) \approx \frac{(-1)^{i+1}}{m} \sum_{k=1}^{m} V_k^{i+1} \exp(-V_k t), t \in (0, \infty)$$

Table 7 VaR, CVaR, and CCVaR results for different copulas and Margins

| | | $\beta = 0.95$ | | | $\beta = 0.99$ | | |
|---|---|---|---|---|---|---|---|
| Copula | Margins | VaR | CVaR | CCVaR | VaR | CVaR | CCVaR |
| Indep. | Normal | 1.2181 | 1.5490 | 5.2625 | 1.7353 | 2.0058 | 6.2988 |
| | Std-t | 1.3482 | 1.7652 | 7.4117 | 2.0092 | 2.4075 | 10.4018 |
| | Ske-t | 1.1027 | 1.4476 | 6.2593 | 1.6625 | 2.0123 | 8.7983 |
| Clayton | Normal | 1.9819 | 2.3482 | 5.2532 | 2.5943 | 2.8642 | 6.4990 |
| | Std-t | 2.1406 | 2.6376 | 7.3899 | 2.9522 | 3.4211 | 10.2903 |
| | Ske-t | 1.7797 | 2.1971 | 6.2407 | 2.4548 | 2.8434 | 8.6836 |
| Frank | Normal | 2.5394 | 2.9858 | 5.2340 | 3.2682 | 3.5758 | 6.2912 |
| | Std-t | 2.7710 | 3.4020 | 7.3489 | 3.8407 | 4.3604 | 10.3683 |
| | Ske-t | 2.2962 | 2.8486 | 6.2057 | 3.2013 | 3.6583 | 8.7690 |
| Gumbel | Normal | 2.7864 | 3.6906 | 5.2371 | 4.2223 | 4.9823 | 6.2774 |
| | Std-t | 3.0446 | 4.4555 | 7.3731 | 5.1980 | 6.7628 | 10.3479 |
| | Ske-t | 2.5184 | 3.7152 | 6.2260 | 4.3158 | 5.7265 | 8.7521 |
| Joe | Normal | 2.8913 | 3.7816 | 5.1140 | 4.3029 | 5.0284 | 6.1605 |
| | Std-t | 3.1185 | 4.5510 | 7.0574 | 5.3501 | 6.9080 | 9.9201 |

|  | Ske-t | 2.5861 | 3.8198 | 5.9571 | 4.5226 | 5.9827 | 8.3894 |
| --- | --- | --- | --- | --- | --- | --- | --- |
| AMH | Normal | 2.0528 | 2.4347 | 5.2512 | 2.6878 | 2.9549 | 6.7231 |
|  | Std-t | 2.2506 | 2.7776 | 7.3850 | 3.0789 | 3.5391 | 12.3847 |
|  | Ske-t | 1.8638 | 2.2978 | 6.2366 | 2.5714 | 2.9540 | 9.8258 |

where $V_k \sim F = \mathcal{LS}^{-1}[\varphi^{(-1)}]$, for $k = 1, \ldots, m$ and $\mathcal{LS}^{-1}[\varphi^{(-1)}]$ represents the inverse of Laplace-Stieljes transform of the inverse generator.

To compare these results, classical VaR using the same estimates for the copula were computed. Following the procedures described in Palaro and Hotta (2004), Patton (2006), Fantazzini (2008), Molina Barreto, Ishimura, and Takaoka (2022) among others; estimation of the Value at Risk at a confidence level $\beta$ can be performed for the total loss return variable $Z = \lambda_1 X_1 + \cdots + \lambda_d X_d$ whose joint distribution is given for a copula $C^{(d)}$ with density $c$ and marginal distributions given by $F_1, \ldots, F_d$, by solving the following equation $\zeta(z^*) = \beta$ where

$$\zeta(z) = P(Z \leq z) = P(\lambda_1 X_1 + \cdots + \lambda_d X_d \leq z)$$

Here $\mathbf{1}_A$ is the indicator function of the subset $A$. The solution $z^*$ is the $\text{VaR}_\beta(Z)$. However, solving this equation is challenging, so using a Monte Carlo approach is preferable. Fortunately, sampling from Archimedean copulas for any dimension is already addressed in Hofert (2008), where efficient algorithms for the copulas utilized in this study are presented. Once sampling from the copula is completed, the empirical $\beta$ quantile can be estimated as the mean for those samples that exceed $\text{VaR}_\beta$ were computed to obtain the corresponding $\text{CVaR}_\beta$ for the portfolio loss. This procedure is performed for every combination of margins and copula.

As stated by Proposition (4), it is observed that $\text{CCVaR}_\beta$ increases as the risk level does so, property that is also shared by the classical $\text{VaR}_\beta$ and $\text{CVaR}_\beta$. Across all copulas, it seems that the Student-T gives the highest values for all risk measures followed by the Skewed-T, evidencing the impact of heavy-tailed distribution over the loss for the portfolio.

According to these numerical results, $\text{CCVaR}_\beta$ surpasses estimates for $\text{CVaR}_\beta$ and for consequence also $\text{VaR}_\beta$, i.e., $\text{CCVaR}_\beta$ is the most conservative risk among the observed measures. for both risk levels. This observation is in line with intuitive idea behind this CCVaR, since expected losses are computed once all the unfavorable events have occurred ( $X \in \mathcal{U}_\beta$ ), unlike the CVaR where expected losses are computed when the weighted sum of the losses has exceeded the VaR at level $\beta$.

Additionally, it is observed that the variation of CCVaR across different copulas is not significant, unlike VaR or CVaR, where copula specification exerts a substantial influence on the final measurement, with the Gumbel and Joe copulas exhibiting the highest values. The lowest CVaR values were those of the independent copula, which contrasts with the CCVaR where its value with the same copula outperforms all copulas when $\beta = 0.95$ considering the same distribution. A very similar thing happens for $\beta = 0.99$, where the value of the CCVaR is higher for the independent copula, except for some distributions with Clayton copula, and AMH.

### 3.2.4 CCVaR Behavior through time

The evolution of CCVaR over time was examined under a fixed dependence structure with varying parameters. To this end, the portfolio described above was employed, and the overall estimation period was partitioned into sub-samples of 1 000 observations each, yielding 359 time windows for parameter estimation of each marginal distribution and the copula (assumed constant across windows). Once parameter estimates were obtained for each window, Value at Risk (VaR), Conditional Value at Risk (CVaR), and CCVaR were calculated using the formula derived in the main result. These results and the actual return were plotted in figures (B4, B5, B6) for levels $\beta = 0.95, 0.99$.

In general, it can be noticed that both CVaR and CCVaR react to periods of high volatility. Also, the order relationship between VaR, CVaR and CCVaR is maintained over time giving the latter as the most conservative risk measure. As mentioned in the previous subsection, the variation of CCVaR does not vary significantly depending on the choice of the copula as opposed to VaR or CVaR. It is also noted how all the measures are increasing with respect to the risk level $\beta$. It seems that the gap that CCVaR exhibits between risk levels is deeper than the rest, implying a notorious impact for more extreme events. Again, the selection of margins seems to be important for the estimation of CCVaR in order to capture the asymmetry and kurtosis for each asset.

# 4 Conclusion and Discussion

Copula-based Conditional Value at Risk has been examined in multivariate settings, and an almost closed-form expression for Archimedean copulas has been derived. The proposed CCVaR is based on an integral transformation involving the generator function and its inverse derivatives, analogous to the Kendall distribution function for Archimedean copulas (Barbe et al., 1996).

Under the specified regularity conditions, CCVaR satisfies the properties of a coherent risk measure. Nevertheless, several open questions remain. For instance, the behavior of CCVaR under increasing dependence parameters in dimensions greater than two has yet to be formally characterized. Although numerical evidence indicates that parameter increases reduce CCVaR—as observed in the bivariate case—a rigorous proof remains elusive. Similarly, the influence of portfolio weights on the measure warrants further investigation.

Numerical results confirm that $\text{CVaR}_\beta(X) < \text{CCVaR}_\beta(X)$, supporting the premise that conditioning on all unfavorable events produces a more conservative risk estimate than conditioning solely on the weighted sum exceeding VaR. These findings underscore the importance of accurate marginal modeling, particularly in the presence of heavy tails and asymmetry.

Future research will examine the behavior of CCVaR in higher dimensions and under alternative dependence structures—such as hierarchical Archimedean and vine copulas—to better capture diverse dependency relationships in diversified portfolios.

# Appendix A Proofs

Proof of Theorem 3. The main result relies in observation made in Barbe et al. (1996); Genest, Nešlehová, and Ziegel (2011); Genest and Rivest (1993); Imlahi and Chakak (1999) regarding the Kendall distribution process and properties of $\varphi$. Assuming that $\varphi$ is $C^d$ class, density for the copula is:

$$dC(u_1, \ldots, u_d) = c(u_1, \ldots, u_d) du_d \cdots du_1$$
$$= \varphi^{-1(d)}(\varphi(u_1) + \cdots + \varphi(u_d)) \prod_{i=1}^{d} \varphi'(u_i) du_d \cdots du_1$$

Now taking $t_d = \varphi^{-1}(\varphi(u_1) + \cdots + \varphi(u_d))$ and considering the change of variable

$$(u_1, \ldots, u_d) \mapsto (u_1, \ldots, u_{d-1}, t_d)$$

and $\varphi'(t_d) dt_d = \varphi'(u_d) du_d$. The projection of the set $\mathcal{U}_\beta^{(d)}$ on $[0,1]^{d-1}$ is the set

$$\mathcal{U}_\beta^{(d-1)} = \{(u_1, \ldots, u_{d-1}) \in [0,1]^{d-1} \mid C^{(d-1)}(u_1, \ldots, u_{d-1}) \geq \beta\}$$

and $\beta \leq t_d \leq \varphi^{-1}(\varphi(u_1) + \cdots + \varphi(u_{d-1})) = C^{(d-1)}(u_1, \ldots, u_{d-1})$. Define $t_{d-1} = \varphi^{-1}(\varphi(u_1) + \cdots + \varphi(u_{d-1}))$ By the symmetry of the variables $u_i$, for $i = 1, \ldots, d$, then for the numerator in $\text{CCVaR}_\beta$, see that:

$$\int \cdots \int_{\mathcal{U}_\beta^{(d)}} \left(\lambda_1 F_{X_1}^{(-1)}(u_1) + \cdots + \lambda_d F_{X_d}^{(-1)}(u_d)\right) dC(u_1, \ldots, u_d)$$
$$= \int \cdots \int_{\mathcal{U}_\beta^{(d-1)}} \left(\sum_{i=1}^{d} \lambda_j F_{X_i}^{(-1)}(u_1)\right) \left(\prod_{i=1}^{d-1} \varphi'(u_i)\right) \left(\int_\beta^{t_{d-1}} \varphi^{-1(d)}(\varphi(t_d)) \varphi'(t_d) dt_d\right) du_{d-1} \cdots du_1$$
$$= \int \cdots \int_{\mathcal{U}_\beta^{(d-1)}} \left(\sum_{i=1}^{d} \lambda_j F_{X_i}^{(-1)}(u_1)\right) \left(\prod_{i=1}^{d-1} \varphi'(u_i)\right) (f_{d-2}(t_{d-1}) - f_{d-2}(\beta)) du_{d-1} \cdots du_1$$

where $f_i(t)$ are defined as in (6 & 7) and using expression (8). Let $h_1(t_{d-1}, \beta) := f_{d-2}(t) - f_{d-2}(\beta)$. Again, considering the change of variable,

$$(u_1, \ldots, u_{d-1}) \mapsto (u_1, \ldots, u_{d-2}, t_{d-1})$$

and making the same consideration for the domain of integration

$$\mathcal{U}_\beta^{(d-2)} = \{(u_1, \ldots, u_{d-2}) \in [0,1]^{d-2} \mid C^{(d-2)}(u_1, \ldots, u_{d-2}) \geq \beta\}$$

and $\beta \leq t_{d-1} \leq \varphi^{-1}(\varphi(u_1) + \cdots + \varphi(u_{d-2})) = C^{(d-2)}(u_1, \ldots, u_{d-2})$. Define $t_{d-2} = \varphi^{-1}(\varphi(u_1) + \cdots + \varphi(u_{d-2}))$, so the last expression is equal to:

$$\int \cdots \int_{\mathcal{U}_\beta^{(d-1)}} \left(\sum_{i=1}^{d} \lambda_j F_{X_i}^{(-1)}(u_1)\right)\left(\prod_{i=1}^{d-1} \varphi'(u_i)\right) h_1(t_{d-1}, \beta) du_{d-1} \cdots du_1$$

$$= \int \cdots \int_{\mathcal{U}_\beta^{(d-1)}} \left(\sum_{i=1}^{d} \lambda_j F_{X_i}^{(-1)}(u_1)\right)\left(\prod_{i=1}^{d-1} \varphi'(u_i)\right) (f_{d-2}(t_{d-1}) - f_{d-2}(\beta)) du_{d-1} \cdots du_1$$

$$= \int \cdots \int_{\mathcal{U}_\beta^{(d-2)}} \left(\sum_{i=1}^{d} \lambda_j F_{X_i}^{(-1)}(u_1)\right)\left(\prod_{i=1}^{d-2} \varphi'(u_i)\right)$$
$$\left(\int_\beta^{t_{d-2}} (f_{d-2}(t_{d-1}) - f_{d-2}(\beta)) \varphi'(t_{d-1}) dt_{d-1}\right) du_{d-2} \cdots du_1$$

$$= \int \cdots \int_{\mathcal{U}_\beta^{(d-2)}} \left(\sum_{i=1}^{d} \lambda_j F_{X_i}^{(-1)}(u_1)\right)\left(\prod_{i=1}^{d-2} \varphi'(u_i)\right)$$

$$= \int \cdots \int_{\mathcal{U}_\beta^{(d-2)}} \left(f_{d-3}(t_{d-2}) - f_{d-3}(\beta) - f_{d-2}(\beta)(\varphi F_{X_i}(t_{d-2}) - \varphi(\beta))\right) du_{d-2} \cdots du_1$$

where $h_2(t, \beta) = f_{d-3}(u) - f_{d-2}(\beta)(\varphi(t) - \varphi(\beta))$ is defined. If this calculation continues, it can be found by induction that in the $j$-th change of variable, the concerned integrand will depend on the function $h_j(t_{d-j}, \beta)$, which has the following form:

$$h_j(t_{d-j}, \beta) := f_{d-j-1}(t_{d-j}) - f_{d-j-1}(\beta) - \sum_{i=1}^{j-1} \frac{f_{d-1-j+i}(\beta)}{i!} (\varphi(t_{d-j}) - \varphi(\beta))^i$$

Multiplying for $\varphi'(t)$ and integrating by parts, the result follows. Process continues until the $(d-2)$-th change of variable is reached, and by re-indexing the sum, the result is obtained.

$$h_{d-2}(t_2, \alpha) = f_1(t_2) - f_1(\beta) - \sum_{i=1}^{d-3} \frac{f_i(\beta)}{i!} (\varphi(t_2) - \varphi(\beta))^i$$

$$\begin{aligned}
\text{CCVaR}_\beta(X) &= \frac{\int_\beta^1 \left(\sum_{i=1}^d \lambda_i F_{X_i}^{(-1)}(t)\right) \varphi'(t) \int_\beta^t \varphi'(t_2) h_{d-2}(t_2, \beta) dt_2 \, dt}{\int_\beta^1 \varphi'(t) \left(\int_\beta^t \varphi'(t_2) h_{d-2}(t_2, \beta) dt_2\right) dt} \\
&= \frac{\int_\beta^1 \left(\sum_{i=1}^d \lambda_i F_{X_i}^{(-1)}(t)\right) \varphi'(t) \int_\beta^t \varphi'(t_2) \left(f_1(t_2) - f_1(\beta) - \sum_{i=1}^{d-3} \frac{f_i(\beta)}{i!} (\varphi(t_2) - \varphi(\beta))^i\right) dt}{\int_\beta^1 \varphi'(t) \int_\beta^t \varphi'(t_2) \left(f_1(t_2) - f_1(\beta) - \sum_{i=1}^{d-3} \frac{f_i(\beta)}{i!} (\varphi(t_2) - \varphi(\beta))^i dt_2\right) dt} \\
&= \frac{\int_\beta^1 \left(\sum_{i=1}^d \lambda_i F_{X_i}^{(-1)}(t)\right) \varphi'(t) \left(f_0(t) - f_0(\beta) - \sum_{i=1}^{d-2} \frac{f_i(\beta)}{i!} [\varphi(t) - \varphi(\beta)]^i\right) dt}{\int_\beta^1 \varphi'(t) \left(f_0(t) - f_0(\beta) - \sum_{i=1}^{d-2} \frac{f_i(\beta)}{i!} [\varphi(t) - \varphi(\beta)]^i\right) dt} \\
&= \frac{\int_\beta^1 \left(\sum_{i=1}^d \lambda_i F_{X_i}^{(-1)}(t)\right) \varphi'(t) h_{d-1}(t, \beta) dt}{1 - K_\varphi^{(d)}(\beta)}
\end{aligned}$$

Proof of Prop. 4. For this observation, note that the $CCVaR_\beta(X)$ can be expressed as a convex of the components of the $\underline{CTE}_\beta(X)$ by Cousin and Di Bernardino (2014). That is,

$$\text{CCVaR}_\beta(X) = \sum_{i=1}^d \lambda_i \underline{CTE}_\beta^i(X)$$

This result is an immediate derivation from their Corollary 2.4.

Proof of Prop. 7. Again, refer to Collorary 2.2 in Cousin and Di Bernardino (2014). If $\theta_1 \leq \theta_2$, the following expression must be satisfied:

$$\frac{1 - t + \varphi_{\theta_2}(t)/\varphi'_{\theta_2}(\beta)}{1 - \beta + \varphi_{\theta_2}(\beta)/\varphi'_{\theta_2}(\beta)} \leq \frac{1 - t + \varphi_{\theta_1}(t)/\varphi'_{\theta_1}(\beta)}{1 - \beta + \varphi_{\theta_1}(\beta)/\varphi'_{\theta_1}(\beta)} \quad \text{for } \beta < t < 1 \quad (A1)$$

where $\varphi_{\theta_1}$ and $\varphi_{\theta_2}$ are generators of copulas $C_{\theta_1}$ and $C_{\theta_2}$ respectively. The previous relationship is indeed satisfied for Clayton and Frank copulas. However, it must be remarked that for Gumbel copula, the above expression seems to not be true for values of $\theta$ close to 1, as can be observed in Figure (4). Moreover, for $\theta < 2$, it was proved that

$\text{CCVaR}_\beta \geq \text{MCVaR}_\beta$ (Molina Barreto and Ishimura, 2023). This means that the $\text{CCVaR}_\beta$ should be decreasing in $\theta \geq 2$. To observe this, call

$$H(\theta) := H(\theta; t, \beta) := \frac{1 - t - \frac{\beta(-\log t)^\theta}{\theta(-\log \beta)^{\theta-1}}}{1 - \beta - \frac{\beta}{\theta}(-\log \beta)}$$

and write $a = -\log t$ and $b = -\log \beta$. Then, the previous definition can be written as

$$H(\theta; a, b) = \frac{1 - e^{-a} - \frac{a^\theta b^{1-\theta} e^{-b}}{\theta}}{1 - e^{-b} - \frac{b e^{-b}}{\theta}}$$

where $0 < a < b$. We need to show that,

$$H_1(\theta) := H_1(\theta; a, b) := H'(\theta) / \frac{a^\theta b^{1-\theta} e^{-a}}{(b + \theta - e^b \theta)^2} \quad (A2)$$

$$= e^a \left(e^b - 1 - \left((e^b - 1)\theta - b\right) \ln \frac{a}{b}\right) - (e^a - 1) e^b \left(\frac{b}{a}\right)^\theta < 0 \quad (A2)$$

Firstly, note that

$$H_1'(2) = (e^a - 1)(e^b - 1)(g(b) - g(a)) \log \frac{a}{b}$$

where $g(a) := \frac{a^2 e^a}{e^a - 1}$ whose derivative is always positive, so $g$ is increasing. Then $H_1'(2) \leq 0$. Next, define

$$H_{12}(b) := \frac{H_1(2)}{e^a(2(e^b - 1) - b)}$$

which will have the same sign of $H_1(2)$, and deriving with respect to $b$:

$$H_{12}'(b) = \frac{(e^a - 1)(e^b - 1)}{a^2 b e^a (2(e^b - 1) - b)^2} (g(a) - g(b))(4(e^b - 1) - b(b + 3))$$

whose value is negative because of $g$. Then $H_{12}(b) \leq H_{12}(a) = 0$. So $H_1(2) \leq 0$. Finally note that $H_1(\theta)$ is concave in $\theta$, because

$$H_1''(\theta) = -\left(\frac{b}{a}\right)^\theta e^b(e^a - 1)\left(\log\frac{b}{a}\right)^2 \leq 0$$

So, $H(\theta)$ is non-increasing in $\theta \geq 2$.

Now, for the Ali-Mikhail-Haq copula, show that the function $H$ is increasing in $\theta \in (0,1)$ and decreasing in $\theta \in (-1,0)$. Write function $H(\theta)$ as:

$$H(\theta) := H(\theta; t, \beta) := \frac{F(\theta)}{G(\theta)}$$

where

$$F(\theta) := F(\theta; t, \beta) := \frac{(t-1)(1-\theta)}{1-(1-\beta)\theta} + \beta\log\frac{1-(1-t)\theta}{t} \text{ and}$$

$$G(\theta) := G(\theta; \beta) := \frac{(\beta-1)(1-\theta)}{1-(1-\beta)\theta} + \beta\log\frac{1-(1-\beta)\theta}{\beta}$$

Note that $G'(t) = (1-\beta)^2 \beta\theta/(1-(1-\beta)\theta)^2$ have the same sign as $\theta$, so $G$ is increasing in $\theta \in (0,1)$ and decreasing in $\theta \in (-1,0)$, and $G(1) = 0$. The derivative of $G(-1)/\beta$ with respect to $\beta$ is always positive, so that $G(-1) = G(-1;\beta)/\beta$ is increasing in $\beta \in (0,1)$ to $G(-1;1)/1 = 0$, so that $G(-1) < 0$. Then,

$$G < 0, G' > 0 \text{ on } (0,1) \ (G' < 0 \text{ on } (-1,0), \text{ resp. })$$

Now, consider the derivative ratio $\rho$:

$$\rho(\theta) := \frac{F'(\theta)}{G'(\theta)} = \frac{(1-t)(1-2\beta + t - (1-\beta)^2\theta)}{(1-\beta)^2(1-(1-t)\theta)}$$

whose derivative $\rho'(\theta) < 0$ and $F(1) = G(1) = 0$. By Prop. 4.1 of Pinelis (2006), the function $H$ is decreasing on $(0,1)$.

# Appendix B Modelling for each margin

Once the negative log-returns have been calculated, the estimation of each margin is proceeded. An ARMA-GARCH model is utilized with three different specifications for the distribution of standardized residuals. By definition, an $\text{ARMA}(p,q) - \text{GARCH}(r,s)$ model is given by:

$$X_t = a_0 + \sum_{i=1}^{p} a_i X_{t-i} + \varepsilon_t + \sum_{j=1}^{q} b_j \varepsilon_{t-j}, \varepsilon_t = z_t \sigma_t$$

$$\sigma_t^2 = c_0 + \sum_{i=1}^{r} c_i \varepsilon_{t-i}^2 + \sum_{j=1}^{s} d_j \sigma_{t-j}^2$$

where $z_t, t = 1, 2, \ldots, T$, is a sequence of independently and identically distributed (i.i.d.) with desired distribution, zero mean and unit variance. Coefficients $c_0, c_1, \ldots, c_r$ and $d_1, \ldots, d_s$ are non-negative and must satisfy the no explosion conditions (Francq & Zakoïan, 2019). Estimation of the parameters by quasi-maximum-likelihood with

Standard Normal, Student-T, and Skewed-T (Fernández & Steel, 1998) distributions for the residuals are presented. Results for each asset of the **G1** are reported in table B1. Values for Akaike's Information Criteria, Ljung-Box Test for squared residual at lags 1 to 10, Kolmogorov-Smirnov test, Chi-Squared Goodness-of-Fit, and AndersonDarling p-values are reported.

After the series is filtered and transformed by their corresponding distribution estimate, a scatter matrix plot is presented for each distribution in figures B1, B2 and B3 where the pair-wise relationship between the variables can be obseved. On the diagonal, the corresponding cdf for the transformed series are plotted and they can be compared with the standard uniform cumulative distribution. As better the fit, the curve should get closer to the uniform one.

A tail dependence test for each pair of variables is performed. This test is described in Reiss and Thomas (1997), where the null hypothesis is tail dependence in the upper tail. Values of the statistic and p-values (inside parenthesis) are shown in the same figures.

| Param. | 6103.T Norm | Std-t | Skd-t | 6113.T Norm | Std-t | Skd-t | 6273.T Norm | Std-t | Skd-t | 6301.T Norm | Std-t | Skd-t | 6302.T Norm | Std-t | Skd-t | 6305.T Norm | Std-t | Skd-t | 6326.T Norm | Std-t | Skd-t |
|---|---|---|---|---|---|---|---|---|---|---|---|---|---|---|---|---|---|---|---|---|---|
| $a_0$ | -0.0487 (0.0526) | -0.0415 (0.0504) | -0.0348 (0.0535) | -0.0824 (0.0440) | -0.0696 (0.0397) | -0.0641 (0.0420) | -0.0693 (0.0503) | -0.0654 (0.0450) | -0.0752 (0.0490) | -0.0923 (0.0510) | -0.0876 (0.0458) | -0.0727 (0.0492) | -0.0632 (0.0519) | -0.0576 (0.0466) | -0.0561 (0.0489) | -0.1194 (0.0636) | -0.0821 (0.0527) | -0.0774 (0.0432) | -0.0275 (0.0471) | -0.0279 (0.0455) | -0.0279 (0.0455) |
| $b_0$ | -0.0113 (0.0289) | -0.0213 (0.0298) | -0.0235 (0.0301) | -0.0061 (0.0271) | -0.0122 (0.0263) | -0.0388 (0.0284) | -0.0031 (0.0290) | -0.011 (0.0247) | 0* | 0.0004 (0.0458) | 0.0331 (0.0288) | 0.0188 (0.0295) | 0.0107 (0.0286) | 0.0162 (0.0319) | 0.0165 (0.0257) | 0.0073 (0.0289) | -0.0118 (0.0267) | -0.0121 |
| $c_0$ | 0.1267 (0.0983) | 0.1209 (0.1080) | 0.1186 (0.1045) | 0.0353 (0.0138) | 0.0126 (0.0088) | 0.0124 (0.0088) | 0.2356 (0.2019) | 0.3144 (0.1664) | 0.3178 (0.2019) | 0.9896 (0.4040) | 0.7603 (0.3092) | 0.7362 (0.3135) | 0.9183 (0.4164) | 0.0353 (0.0136) | 0.9168 (0.3140) | 0.5773 (0.2698) | 0.5411 (0.2730) | 0.1031 (0.1507) | 0.0687 (0.0191) | 0.0853 (0.0537) |
| $c_1$ | 0.0567 (0.0338) | 0.0494 (0.0371) | 0.0495 (0.0368) | 0.0222 (0.0050) | 0.0224 (0.0088) | 0.0223 (0.0037) | 0.0381 (0.0208) | 0.0522 (0.0197) | 0.0522 (0.0195) | 0.1211 (0.0451) | 0.1422 (0.0430) | 0.1393 (0.0434) | 0.159 (0.0523) | 0.0244 (0.0065) | 0.069 (0.0356) | 0.0793 (0.0350) | 0.0773 (0.0363) | 0.0317 (0.0071) | 0.0207 (0.0191) | 0.0303 (0.0156) |
| $d_1$ | 0.9102 (0.0597) | 0.9191 (0.0597) | 0.9196 (0.0585) | 0.9663 (0.0034) | 0.9737 (0.0011) | 0.9738 (0.0010) | 0.9005 (0.0090) | 0.8671 (0.0562) | 0.8663 (0.0561) | 0.5958 (0.1302) | 0.643 (0.1105) | 0.6525 (0.1136) | 0.5859 (0.1480) | 0.9654 (0.0031) | 0.9639 (0.0542) | 0.7404 (0.0830) | 0.7965 (0.0867) | 0.937 (0.0741) | 0.9588 (0.0043) | 0.9439 (0.0281) |
| $\zeta$ | | | 1.0185 (0.0414) | | | 1.0144 (0.0374) | | | 0.9758 (0.0378) | | | 1.0318 (0.0380) | | | 1.0883 (0.0390) | | | 1.0615 (0.0355) | | | 1.1447 (0.0471) |
| $\nu$ | | 8.2169 (1.6705) | 8.2018 (1.6496) | | 6.5263 (1.0007) | 6.5103 (0.9952) | | 7.0509 (1.2029) | 7.026 (1.1935) | | 5.3818 (0.7885) | 5.382 (0.7877) | | 6.5931 (1.3380) | 6.6576 (1.3757) | | 4.893 (0.7225) | 4.979 (0.7402) | | 6.497 (1.0632) | 6.7519 (1.1030) |
| AIC | 4.1551 | 4.1361 | 4.1374 | 3.9806 | 3.9385 | 3.9398 | 4.1829 | 4.1474 | 4.1486 | 4.0569 | 3.9664 | 3.9674 | 4.0706 | 4.0305 | 4.0282 | 4.3878 | 4.2459 | 4.2454 | 4.0549 | 3.9891 | 3.9821 |
| $Q^2(1)$ | 0.0164 | 0.0057 | 0.006 | 0.0592 | 0.109 | 0.1106 | 0.7157 | 0.747 | 0.7481 | 0.8467 | 0.5899 | 0.6035 | 0.8813 | 0.0034 | 0.6797 | 0.6128 | 0.6167 | 0.4204 | 0.5883 | 1.0632 | 0.5946 |
| $Q^2(10)$ | 0.2295 | 0.0665 | 0.0674 | 0.1451 | 0.1999 | 0.2037 | 0.8172 | 0.9951 | 0.9875 | 0.9928 | 0.4416 | 0.519 | 1 | 1 | 0.9401 | 0.9627 | 0.9689 | 0.9689 | 0.9689 | 0.9689 | 0.9689 |
| KS | 0* | 0* | 0* | 0* | 0* | 0* | 0* | 0* | 0* | 0* | 0* | 0* | 0* | 0* | 0* | 0* | 0* | 0* | 0* | 0* | 0* |
| $\chi^2$ | 0.0126 | 0.4213 | 0.3761 | 0.0065 | 0.8595 | 0.811 | 0.0187 | 0.9359 | 0.9731 | 0.9156 | 0.823 | 0.288 | 0.0012 | 0.6022 | 0.7708 | 0.0097 | 0.8903 | 0.5299 | 0.5299 | 0.5299 | 0.9511 |
| AD | 0.0621 | 0.0013 | 0.9773 | 0.0272 | 0* | 0.9961 | 0.0357 | 0.0002 | 0.998 | 0.9408 | 0* | 0* | 0.0215 | 0.9967 | 0* | 0.0047 | 0.778 | 0* | 0* | 0* | 0.9879 |

*Note:* The values in parentheses indicate standard errors.

*Abbreviations:* Norm = Normal, Std-t = Student's t, Skd-t = Skewed-t, AIC = Akaike Information Criterion, $Q^2(1)$ = 1st order squared auto-correlation, $Q^2(10)$ = 10th order squared autocorrelation, (P-values) KS = Kolmogorov-Smirnov test, $\chi^2$ = Chi-squared test, AD = Anderson-Darling test.

**Table B1** ARMA-GARCH Parameter estimates for marginal modeling of negative log-returns for daily price of **G1**

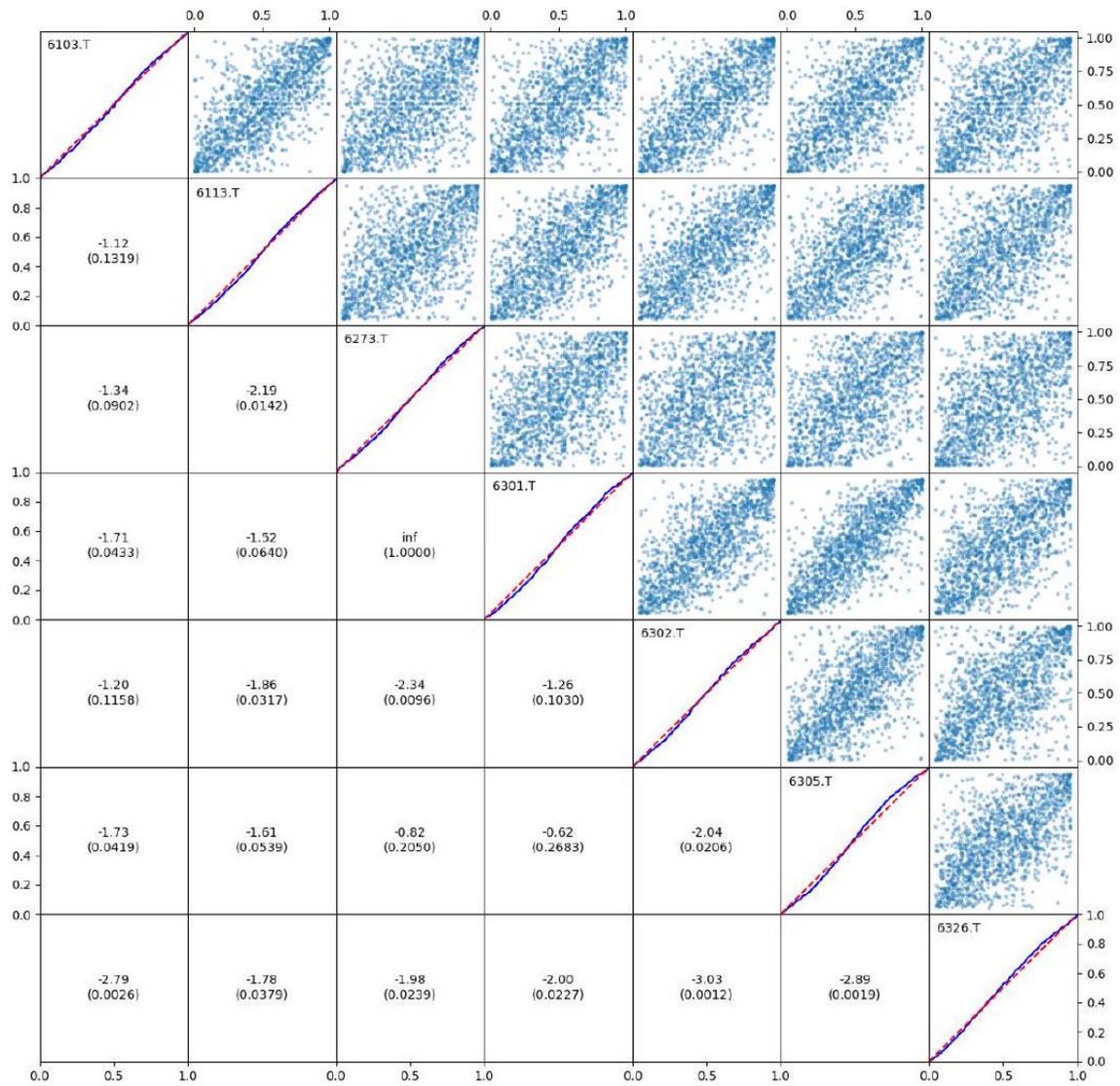

Fig. B1 Scatter plot matrix for standardized transformed margins with normal distributed innovations for assets of G1. Plots on diagonal show the empirical CDF and the standard uniform CDF.

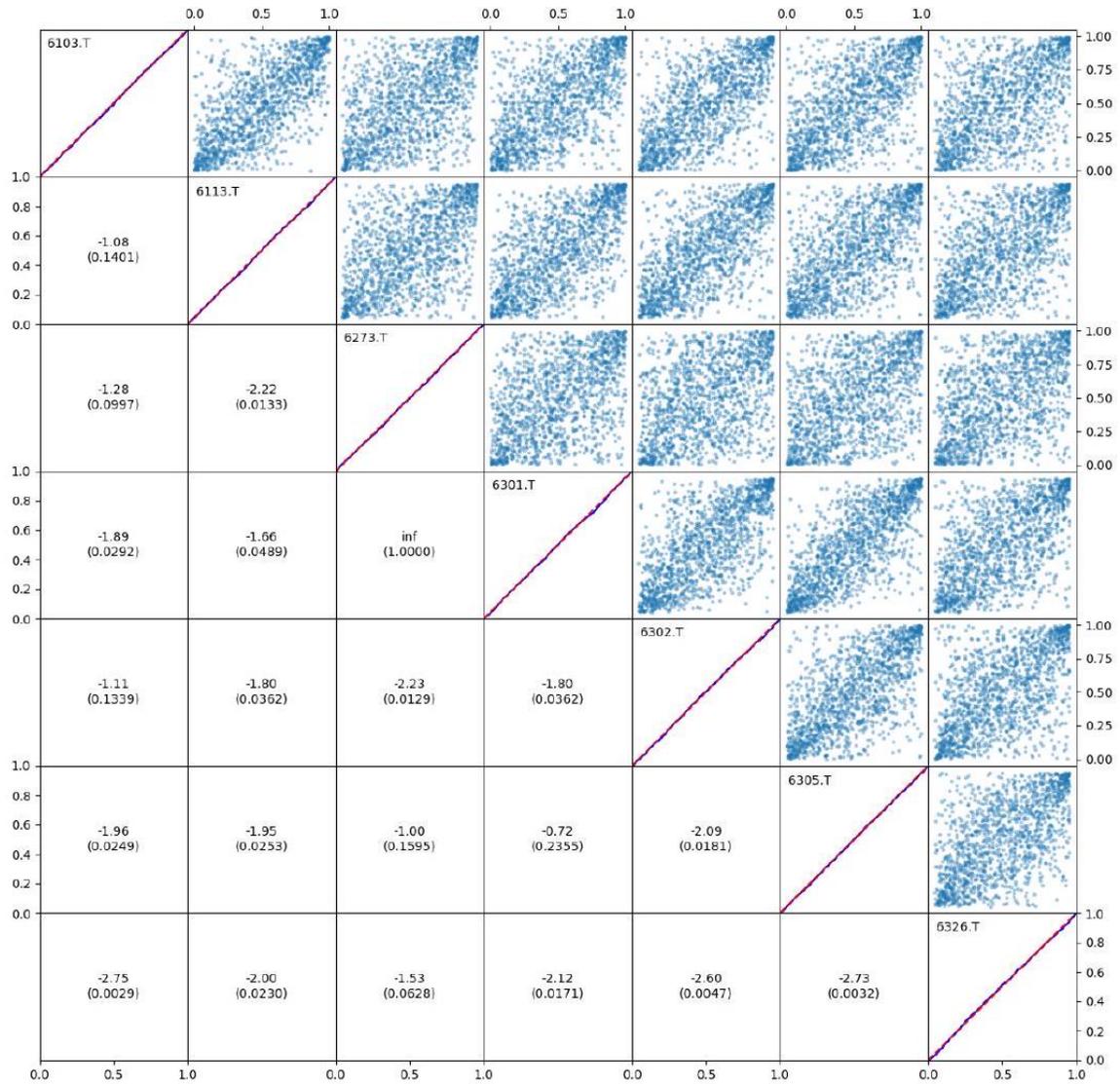

Fig. B2 Scatter plot matrix for standardized transformed margins with Student-t distributed innovations for assets of G1. Plots on diagonal show the empirical CDF and the standard uniform CDF.

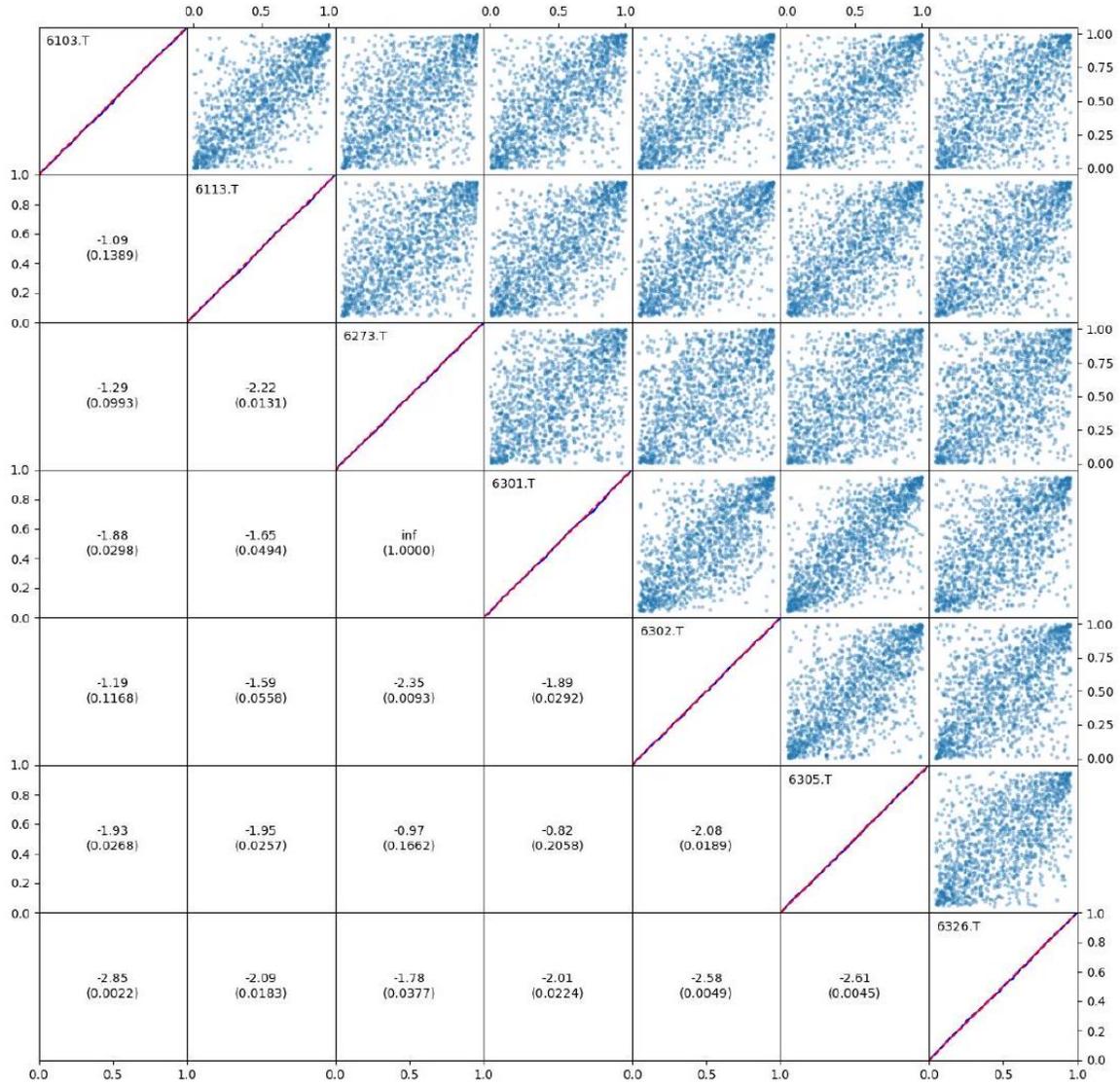

Fig. B3 Scatter plot matrix for standardized transformed margins with Skewed-t distributed innovations for assets of G1. Plots on diagonal show the empirical CDF and the standard uniform CDF.

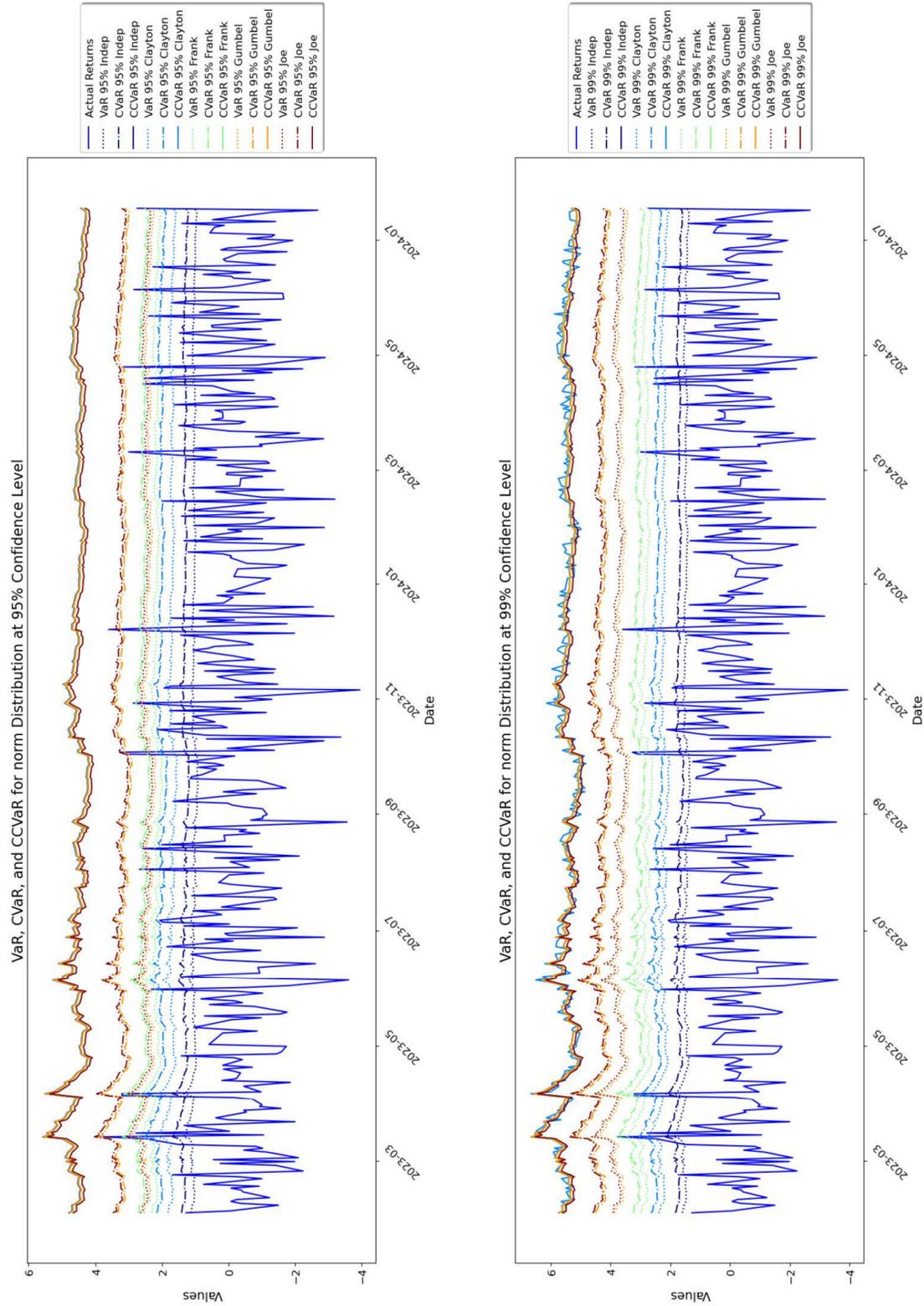

Fig. B4 1-day ahead VaR, CVaR and CCVaR forecasts for portofolio conformed with assets of G1 of equal weights and modeled with standard normal innovations.

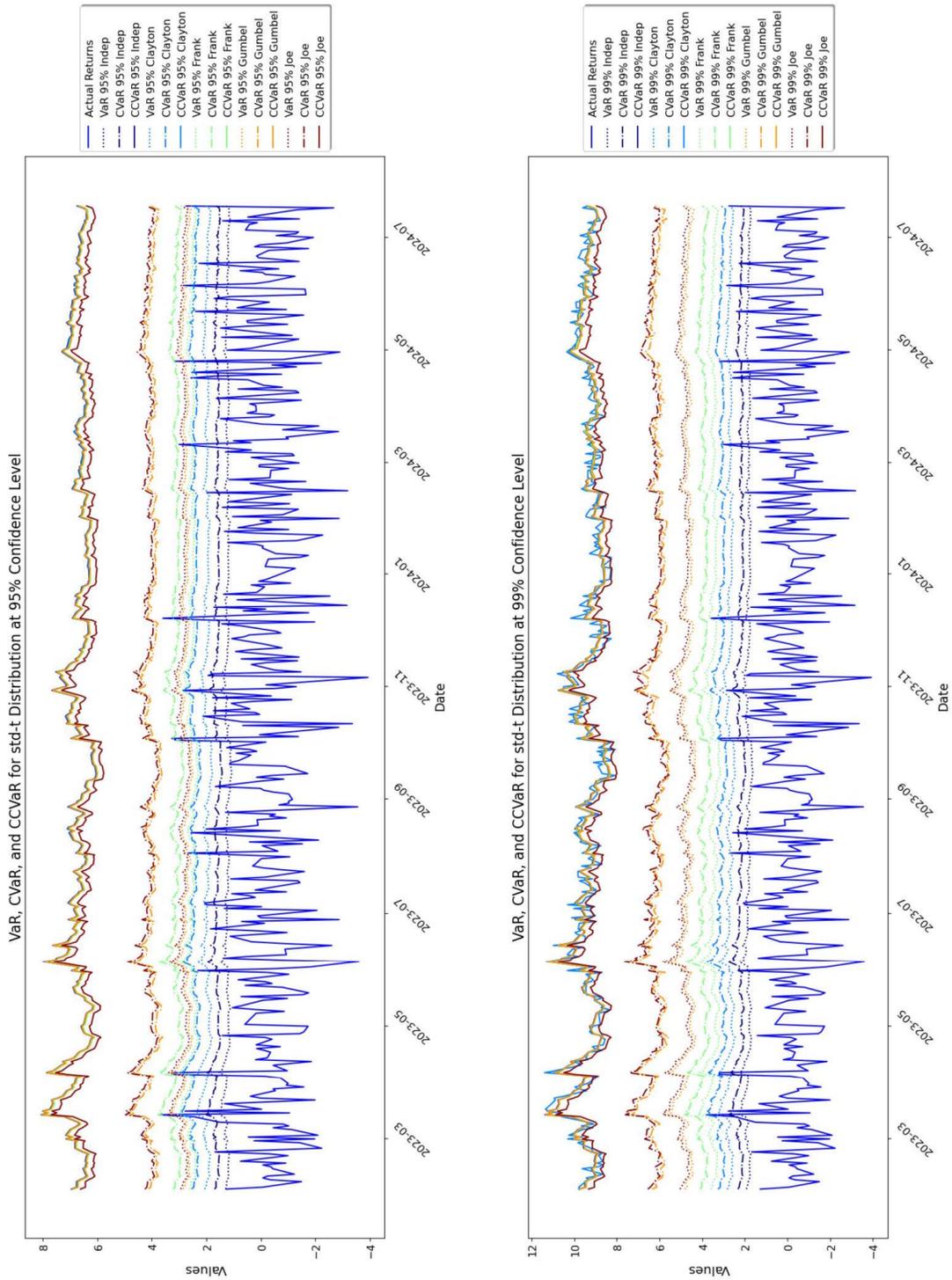

Fig. B5 1-day ahead VaR, CVaR and CCVaR forecasts for portofolio conformed with assets of G1 of equal weights and modeled with Student-t innovations.

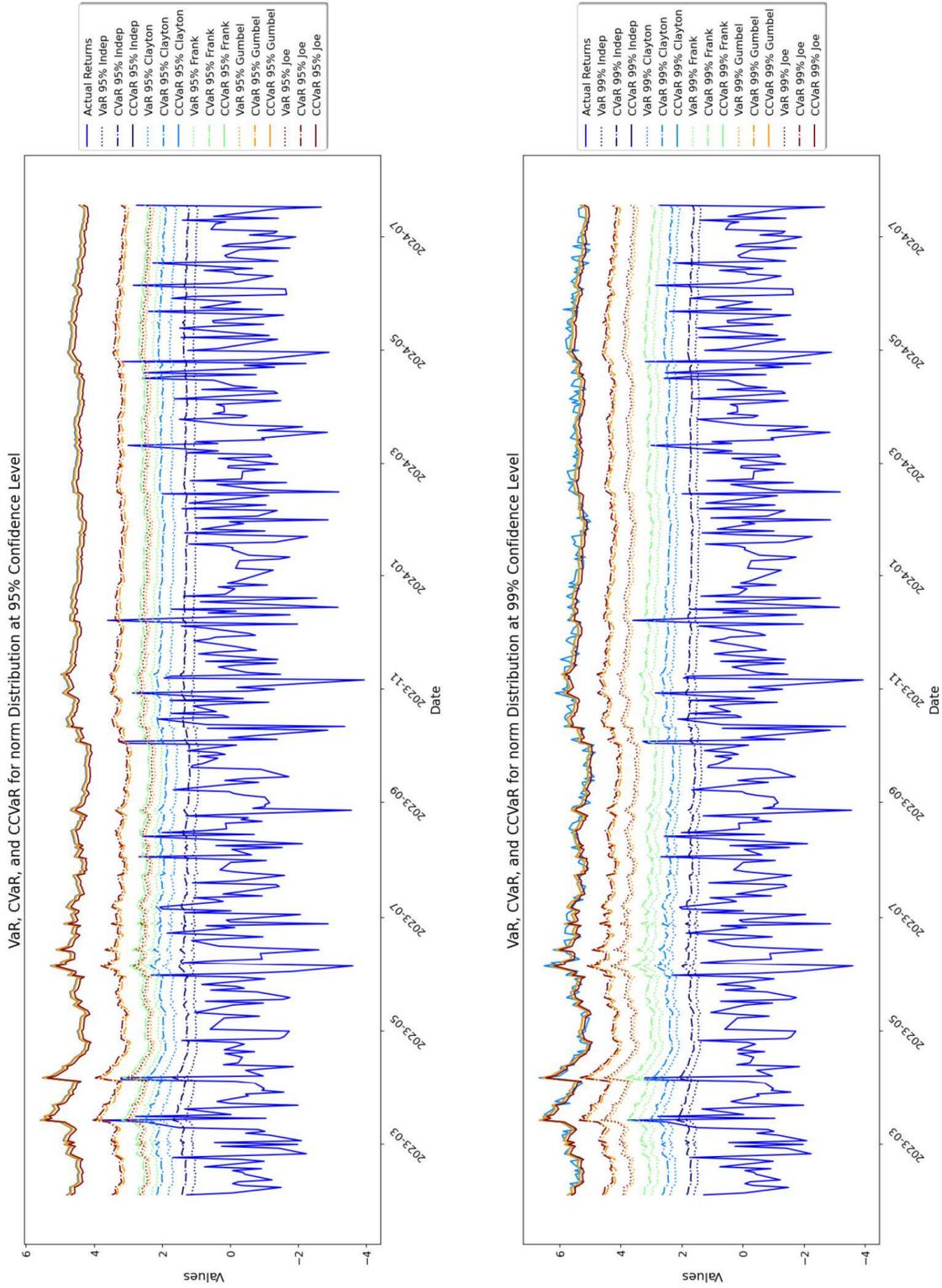

Fig. B6 1-day ahead VaR, CVaR and CCVaR forecasts for portofolio conformed with assets of G1 of equal weights and modeled with Skewed-t innovations

# Data Availability Statement

All the data and Python and R codes for results can be downloaded from the Author's Github repository.

# Statements and Declarations

No potential conflict of interest was reported by the author(s).

# Funding

This work was supported by Keio University Academic Development Funds for Individual Research with number M02JA25062.